\documentclass[]{pasj02} 
\usepackage{natbib} 

\usepackage{url}
\usepackage[usenames,dvipsnames]{xcolor}

\jyear{2024}
\Received{}
\Accepted{}

\begin{document} 

\title{Bulk and turbulent gas motions in the interacting galaxy cluster Abell 3395 South observed with XRISM}

\author{
 Naomi \textsc{Ota},\altaffilmark{1,2}\altemailmark\orcid{0000-0002-2784-3652} \email{naomi@cc.nara-wu.ac.jp} 
 Angie \textsc{Veronica},\altaffilmark{2}\orcid{0000-0002-8635-0030}
 Jakob \textsc{Dietl},\altaffilmark{2,3}\orcid{0009-0001-6898-0004}
 Anri \textsc{Yanagawa},\altaffilmark{1}\orcid{0009-0009-3388-2509}
 Thomas H. \textsc{Reiprich},\altaffilmark{2}\orcid{0000-0003-2047-2884}
 Veronica \textsc{Biffi},\altaffilmark{4,5}\orcid{0000-0001-9260-3826}
 Klaus \textsc{Dolag},\altaffilmark{6,7}\orcid{0000-0003-1750-286X}
 Marcus \textsc{Br\"uggen},\altaffilmark{8}\orcid{0000-0002-3369-7735}
 Esra \textsc{Bulbul}, \altaffilmark{9}\orcid{0000-0002-7619-5399}
 Florian \textsc{Pacaud},\altaffilmark{2}\orcid{0000-0002-6622-4555}
 Yoshiki \textsc{Toba}\altaffilmark{10,1,11,12}\orcid{0000-0002-3531-7863}
}
\altaffiltext{1}{Department of Physics, Nara Women’s University, Kitauoyanishi-machi, Nara, Nara 630-8506, Japan}
\altaffiltext{2}{Argelander-Institut f\"{u}r Astronomie (AIfA), Universit\"{a}t Bonn, Auf dem H\"{u}gel 71, 53121 Bonn, Germany}
\altaffiltext{3}{Max-Planck-Institut f\"{u}r Radioastronomie (MPIfR), Auf dem H\"{u}gel 69, 53121 Bonn, Germany}
\altaffiltext{4}{INAF - Osservatorio Astronomico di Trieste, Via Tiepolo 11, I-34143 Trieste, Italy}
\altaffiltext{5}{IFPU - Institute for Fundamental Physics of the Universe, Via Beirut 2, I-34014 Trieste, Italy}
\altaffiltext{6}{Universit\"{a}ts-Sternwarte, Fakult\"{a}t f\"{u}r Physik, Ludwig-Maximilians-Universit\"{a}t M\"{u}nchen, Scheinerstr.1, 81679 M\"{u}nchen, Germany}
\altaffiltext{7}{Max-Planck-Institut f\"{u}r Astrophysik, Karl-Schwarzschild-Stra\ss e 1, 85741 Garching, Germany}
\altaffiltext{8}{University of Hamburg, Hamburger Sternwarte, Gojenbergsweg 112, 21029 Hamburg, Germany}
\altaffiltext{9}{Max-Planck-Institut f\"{u}r extraterrestrische Physik, Giessenbachstra\ss e 1, 85748 Garching, Germany}
\altaffiltext{10}{Department of Physical Sciences, Ritsumeikan University, 1-1-1 Noji-higashi, Kusatsu, Shiga 525-8577, Japan}
\altaffiltext{11}{Academia Sinica Institute of Astronomy and Astrophysics, 11F of Astronomy-Mathematics Building, AS/NTU, No.1, Section 4, Roosevelt Road, Taipei 10617, Taiwan}
\altaffiltext{12}{Research Center for Space and Cosmic Evolution, Ehime University, 2-5 Bunkyo-cho, Matsuyama, Ehime 790-8577, Japan}



\maketitle

\begin{abstract}
We investigate the gas motions in the core region of the Abell~3395 South subcluster (A3395S) using high-resolution X-ray spectroscopy with XRISM/Resolve. By analyzing the Fe~XXV He$\alpha$ emission line, we directly measure the line-of-sight bulk and turbulent velocities of the intracluster medium. 
We find that the one-dimensional turbulent velocity is $124\pm21~{\rm km\,s^{-1}}$, corresponding to a subsonic Mach number of $0.19\pm0.03$, while a finite line-of-sight bulk velocity of $263\pm23~{\rm km\,s^{-1}}$ is detected.
The coexistence of low turbulence and finite bulk motion suggests that A3395S has not yet reached a dynamically relaxed state. These results are consistent with the non-detection of a radio halo in A3395S, suggesting that turbulent particle reacceleration is currently inefficient in the cluster core. This study demonstrates that high-resolution X-ray spectroscopy with XRISM provides a powerful means to directly constrain intracluster medium dynamics in merging galaxy clusters, and it provides a reference for future comparative studies of A3395N and A3391 within the same large-scale structure.
\end{abstract}


\section{Introduction}\label{sec:intro}
Galaxy clusters are the most massive gravitationally bound objects in the Universe, with most of their baryonic mass residing in the intracluster medium (ICM), a hot plasma with temperatures of several keV (e.g., \cite{Boehringer10}). Gas motions in the ICM provide key insights into cluster assembly processes, including mergers, active galactic nucleus (AGN) feedback, and gas accretion along large-scale filaments, and they influence energy dissipation, metal transport, and non-thermal phenomena such as diffuse radio emission.

Direct measurements of ICM bulk and turbulent motions were long limited by the spectral resolution of X-ray instruments. Prior to the advent of microcalorimeter spectroscopy, constraints were obtained from emission-line measurements with CCD and grating instruments as well as from indirect methods based on surface-brightness and pressure fluctuations, although with limited precision (e.g., \cite{Ota12,Sanders23,Zhuravleva18}). High-resolution X-ray spectroscopy now enables accurate Doppler measurements of emission-line centroids and widths. Numerical simulations have demonstrated that Fe-line shifts and broadening provide robust probes of ICM kinematics under realistic observational conditions \citep{Dolag05,Ota18,Biffi22b}. 
This capability was first demonstrated by Hitomi in the Perseus cluster, revealing subsonic turbulence in the core ($\sim160~{\rm km\,s^{-1}}$) \citep{Hitomi16,Hitomi18}. The observed velocity dispersions primarily reflect motions on spatial scales of a few $10$~kpc, set by the line-of-sight effective length of the emission. Recent XRISM observations have further shown that the velocity field is scale-dependent, with different physical processes operating on different spatial scales, including small-scale AGN-driven motions and larger-scale merger-driven motions \citep{XRISM26a}. This implies that the inferred level of turbulence depends on the spatial scale probed by the observation.

The X-ray Imaging and Spectroscopy Mission (XRISM; \cite{Tashiro25}), equipped with the Resolve microcalorimeter \citep{Ishisaki18}, enables systematic measurements of ICM gas motions in nearby clusters. 
Early XRISM results have revealed a wide diversity of dynamical states and highlighted tensions with numerical simulations, particularly in cluster cores where simulations often predict higher velocity dispersions than observed on average, while some individual systems remain consistent with the simulations \citep{XRISM25c}.
These findings motivate interpreting individual XRISM measurements in the context of each cluster's assembly history and environment.

Cosmological simulations predict that ICM dynamics depend on a cluster's connectivity to the surrounding large-scale structure: clusters embedded in filaments are expected to exhibit enhanced bulk motions and anisotropic velocity fields driven by continuous accretion and repeated interactions, while a substantial fraction of the kinetic energy may remain in coherent motions rather than being fully dissipated into small-scale turbulence and heat \citep{Zinger16,Gouin21}. Observational tests of these predictions require measurements of both bulk and turbulent gas motions in well-characterized filamentary environments.

The Abell~3391--3395 system is a nearby ($z \simeq 0.05$) galaxy cluster complex embedded in a prominent large-scale filament, consisting of Abell~3395 South (A3395S), Abell~3395 North (A3395N), and Abell~3391 (A3391). Previous X-ray observations revealed disturbed morphologies and temperature structures in A3395S and A3395N \citep{Forman81,Markevitch98,Tittley01,Lakhchaura11}. On larger scales, diffuse X-ray emission bridging A3391 and A3395 has been detected \citep{Planck13,Sugawara17,Alvarez18}, and Spectrum-Roentgen-Gamma (SRG)/eROSITA \citep{Predehl21} uncovered a $\sim$15~Mpc warm--hot filament and a bridge-like X-ray structure between A3395S and A3395N \citep{Reiprich21,2022A&A...661A..46V,Biffi22a}, establishing this system as part of the cosmic web (figure~\ref{fig:eR_image}a). ASKAP/EMU observations further show that A3395S hosts a central FR~I radio galaxy but no diffuse radio halo or bridge emission \citep{Brueggen21}, providing additional constraints on turbulence and particle acceleration.

In this paper, we present XRISM/Resolve observations of A3395S and directly measure the line-of-sight bulk and turbulent velocities of the ICM from the Fe~XXV He$\alpha$ line. We assess the dynamical state of the A3395S core in the interacting A3395S--A3395N system and compare the ICM redshift with optical galaxy redshifts to test whether the ICM is dynamically relaxed or retains signatures of ongoing interaction. 

This paper is organized as follows: Section~\ref{sec:obs} describes the observations and data reduction. Section~\ref{sec:analysis} presents the spectral analysis and results. In Section~\ref{sec:discuss} we discuss implications for AGN feedback, merger geometry, and comparisons with numerical simulations. Section~\ref{sec:summary} summarizes our conclusions. Throughout this paper, we adopt a flat $\Lambda$CDM cosmology with $\Omega_{\rm m0}=0.3$, $\Omega_{\Lambda}=0.7$, and $h=0.7$. At $z=0.0525$, $1\arcmin$ corresponds to 61~kpc. All redshifts and velocities are corrected to the Solar System barycenter, and elemental abundances are given relative to the proto-solar values of \citet{Lodders09}.

\section{Observations and data reduction}\label{sec:obs}
\subsection{XRISM observation}
Abell~3395 South (hereafter A3395S) was observed with XRISM as part of Announcement of Opportunity~1 (AO1). The observation was carried out using both the Resolve microcalorimeter \citep{Ishisaki18} and the Xtend X-ray imaging spectrometer \citep{Noda25,Uchida25}. Resolve provides high energy resolution required for precise measurements of emission-line centroids and widths from the ICM, while Xtend offers complementary imaging information over a wider field of view.

\begin{figure*}[htb]
 \begin{center}
\includegraphics[width=0.4\linewidth,angle=0]{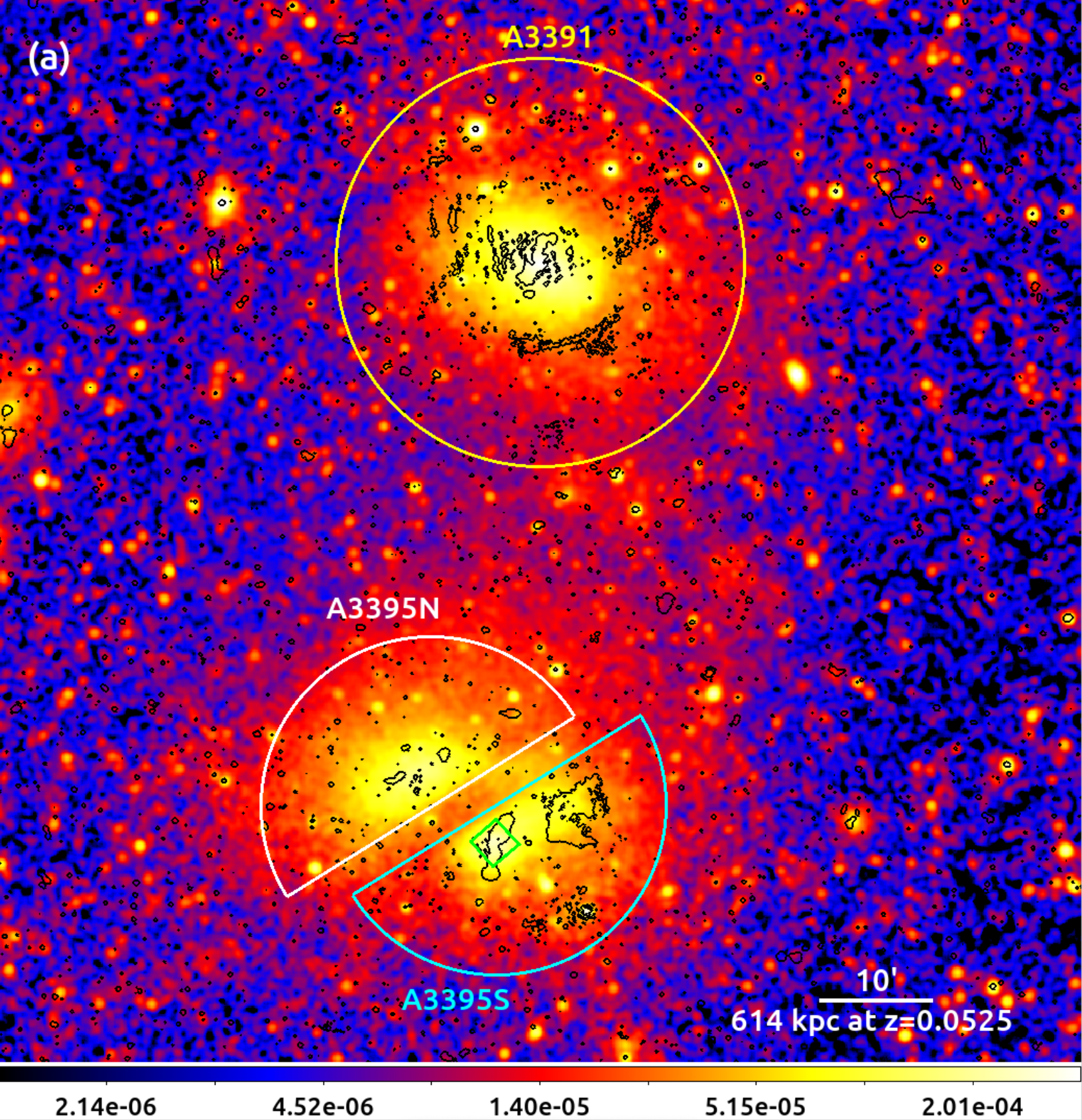}
\includegraphics[width=0.45\linewidth,angle=0]{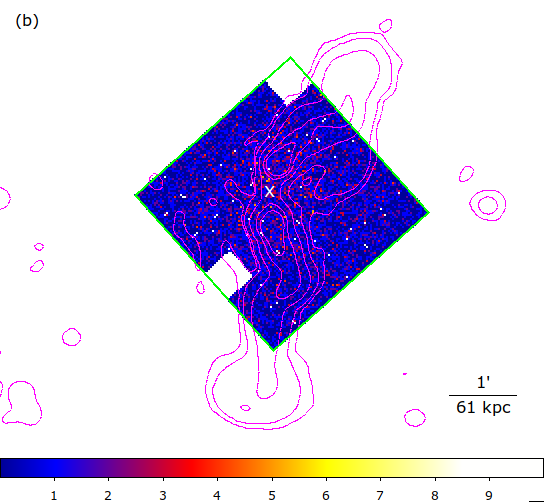}
 \end{center}
 \caption{(a) eROSITA X-ray image of the Abell~3391/95 system in the 0.3--2.0~keV band, particle background-subtracted, corrected for Galactic absorption, exposure-corrected, and smoothed with a 32\arcsec\ Gaussian kernel. The XRISM/Resolve field of view is overlaid as a green box. A single ASKAP/EMU radio contour at $0.0001~{\rm Jy\,beam^{-1}}$ is shown in black \citep{Brueggen21}. Some of the radio structure around A3391 is due to imaging artefacts. The regions used for the optical galaxy redshift analysis in Section~\ref{subsec:optical_data} are indicated: a yellow circle for A3391, and white and cyan semicircles for A3395N and A3395S, respectively. 
(b) XRISM/Resolve X-ray image of A3395S, with the same field of view outlined in green and ASKAP/EMU radio contours overlaid. The position of the brightest cluster galaxy (BCG) is marked with a cross. The radio contour levels in panel~(b) are $0.0001, 0.001, 0.01, 0.05,$ and $0.1~{\rm Jy\,beam^{-1}}$. 
{Alt text: Two-panel X-ray images illustrating the large-scale environment and the central region of the Abell 3391/95 system. Panel (a) shows the three clusters and the intercluster regions. Panel (b) focuses on the southern subcluster A3395S, where radio emission associated with the central radio galaxy traces extended jet structures.}}\label{fig:eR_image}
\end{figure*}

The XRISM observation of A3395S was performed as a single pointing, with the cluster center placed near the center of the Resolve field of view. Details of the observation, including the observation ID, date, and total exposure time, are summarized in table~\ref{tab:observation_log}. In this paper, measurements of gas velocities are based on the Resolve data, while Xtend data are used to assess background components. Figure~\ref{fig:eR_image}a shows the eROSITA X-ray image of A3395S with the Resolve field of view overlaid as a green box.

\begin{table*}[htb]
 \tbl{Observation Log for A3395S}{
    \begin{tabular}{lllllll}\hline\hline
        Object & OBSID &  Coordinates$^{a}$ & DATE  & Exposure$^{b}$ & Fe XXV He$\alpha$$^{c}$ & Fe XXVI Ly$\alpha$$^{c}$ \\
        \hline
        Abell3395\_South & 201020010 & 96.7024, -54.546 & 2025-02-03 & 266.4 & 224 & 28 \\
        \hline
    \end{tabular}}    \label{tab:observation_log}
    \begin{tabnote}
        $^{\mathrm{a}}$ Pointing coordinates. $^{\mathrm{b}}$ Net exposure time after data screening in kilo seconds.  $^{\mathrm{c}}$ Iron line counts without continuum component.
    \end{tabnote}
\end{table*}

\subsection{Data processing and screening}
Data processing and analysis were performed using the standard XRISM pipeline implemented in HEASoft~6.36, together with the XRISM calibration database version~12 (20250915). Event files were reprocessed following the standard screening criteria (The XRISM Data Reduction Guide). Inspection of the light curve confirmed that the count rate was stable throughout the observation, with no significant temporal variations. Only high-primary grade Resolve events were used for the spectral analysis.

High-resolution Resolve event data were used for the spectral analysis. The non--X-ray background (NXB) was estimated following the XRISM Resolve NXB database procedure described in the XRISM analysis documentation\footnote{\url{https://heasarc.gsfc.nasa.gov/docs/xrism/analysis/nxb/resolve_nxb_db.html}}.
NXB spectra were extracted from the dedicated NXB database corresponding to the observation conditions. The extracted NXB spectra were fitted with the standard NXB spectral model, and the best-fit model was used to determine the NXB normalization applied in the subsequent spectral analysis.

The cosmic X-ray background (CXB) was modeled using Xtend data, following the spectral components summarized in table~3 of \citet{Veronica24}. The CXB model consists of two Galactic thermal components described by \texttt{apec} models and an extragalactic power-law component. CXB spectra were extracted from circular regions located in the outer part of the Xtend field of view, outside the $R_{500}$ radii of both A3395S and A3395N, where contamination from cluster emission is negligible. The extraction regions were centered at (RA, Dec) = (96.73905,-55.0057) with radius of 4.5~arcmin. The spectra were fitted assuming a spatially uniform surface brightness, using ancillary response files generated for a uniform sky. During the CXB fitting, the Xtend NXB was included as a fixed template model provided by the XRISM team\footnote{\url{https://heasarc.gsfc.nasa.gov/docs/xrism/analysis/nxb/nxb_spectral_models.html}}. The best-fit normalizations of the CXB components were then rescaled according to the solid angle of the Resolve field of view and included as an additional background component in the Resolve spectral analysis. In the Resolve fits, all CXB model parameters were kept fixed to the values determined from the Xtend analysis.

The energy scale and spectral resolution of Resolve were assessed based on the publicly available gain report\footnote{\url{https://heasarc.gsfc.nasa.gov/FTP/xrism/postlaunch/gainreports/2/201020010_resolve_energy_scale_report.pdf}}. According to this report, pixel~1 and pixel~16 exhibit slightly degraded energy resolution compared to the other pixels, with full width at half maximum values of approximately 5~eV. To evaluate the impact of these pixels on the spectral analysis, we performed fits both including and excluding pixel~1 and pixel~16. We confirmed that the derived temperatures, redshifts, bulk velocities, and turbulent velocities are consistent within statistical uncertainties. In the following analysis, we use all science pixels except pixel~27 and the calibration pixel (shown in white in figure~\ref{fig:eR_image}(b)).

Spectra were extracted from a region centered on the cluster core. The Resolve field of view corresponds to a radius of about $1.5\arcmin$, that is $\sim 0.065R_{500}$ for A3395S ($R_{500}=22'.94=1.39$~Mpc; \cite{Reiprich21}), and thus primarily probes the central core region of the cluster. XRISM measurements are dominated by the dense core emission,
but remain projected quantities, implying an intrinsic uncertainty related to line-of-sight integration. Possible contamination from point sources was assessed using existing high-angular-resolution X-ray observations and was found to be negligible. The Resolve detector response matrices (RMFs) were generated using the standard XRISM tools. The ancillary response files (ARFs) were constructed assuming an extended source, using a Chandra image within a circular region of radius $3'$ as input.

Spectral fitting was performed using an absorbed thermal plasma model, \texttt{tbabs}~$\times$~\texttt{bapec} \citep{Smith01,Foster12}. The Galactic hydrogen column density was fixed at $7.8\times10^{20}~{\rm cm^{-2}}$ \citep{HI4PI16}. The NXB and CXB components described above were included as background models in the spectral fitting. Spectral fitting and error estimation were performed using the C-statistic on unbinned Resolve spectra.  All spectral analyses were carried out with XSPEC version~12.15.1 \citep{Arnaud96}, adopting AtomDB version~3.1.3. We also verified that binning the spectra to ensure at least one count per bin yields consistent results within the statistical uncertainties.

\subsection{Optical data and galaxy redshifts}\label{subsec:optical_data}

Optical spectroscopic redshift information was used to characterize the galaxy component of the Abell~3391--3395 system and to provide a reference for comparison with the ICM redshift measured by XRISM. Heliocentric galaxy redshifts were collected from the NASA/IPAC Extragalactic Database (NED), selecting galaxies located in the vicinity of A3395S, A3395N, and A3391.

The A3395 system consists of two closely separated components, A3395 South (A3395S) and A3395 North (A3395N), which are close both in projected position and in redshift space. To associate galaxies with each component, we applied spatial selection criteria based on cluster-centric apertures chosen to represent each subcluster and examined the resulting redshift distributions. Because the derived mean redshifts depend on the adopted galaxy selection, the scatter among results obtained with different selection methods is treated as an estimate of the systematic uncertainty. The regions used for the optical galaxy redshift analysis are indicated in figure~\ref{fig:eR_image}.

For A3391, the mean cluster redshift was evaluated using only galaxies located within $R_{500}$. This choice was made to minimize contamination from galaxies in the bridge region between A3391 and A3395, which tend to have systematically lower redshifts and can bias the cluster mean when galaxies at larger radii are included.

\begin{figure}[htb]
 \begin{center}
  \includegraphics[width=0.9\linewidth]{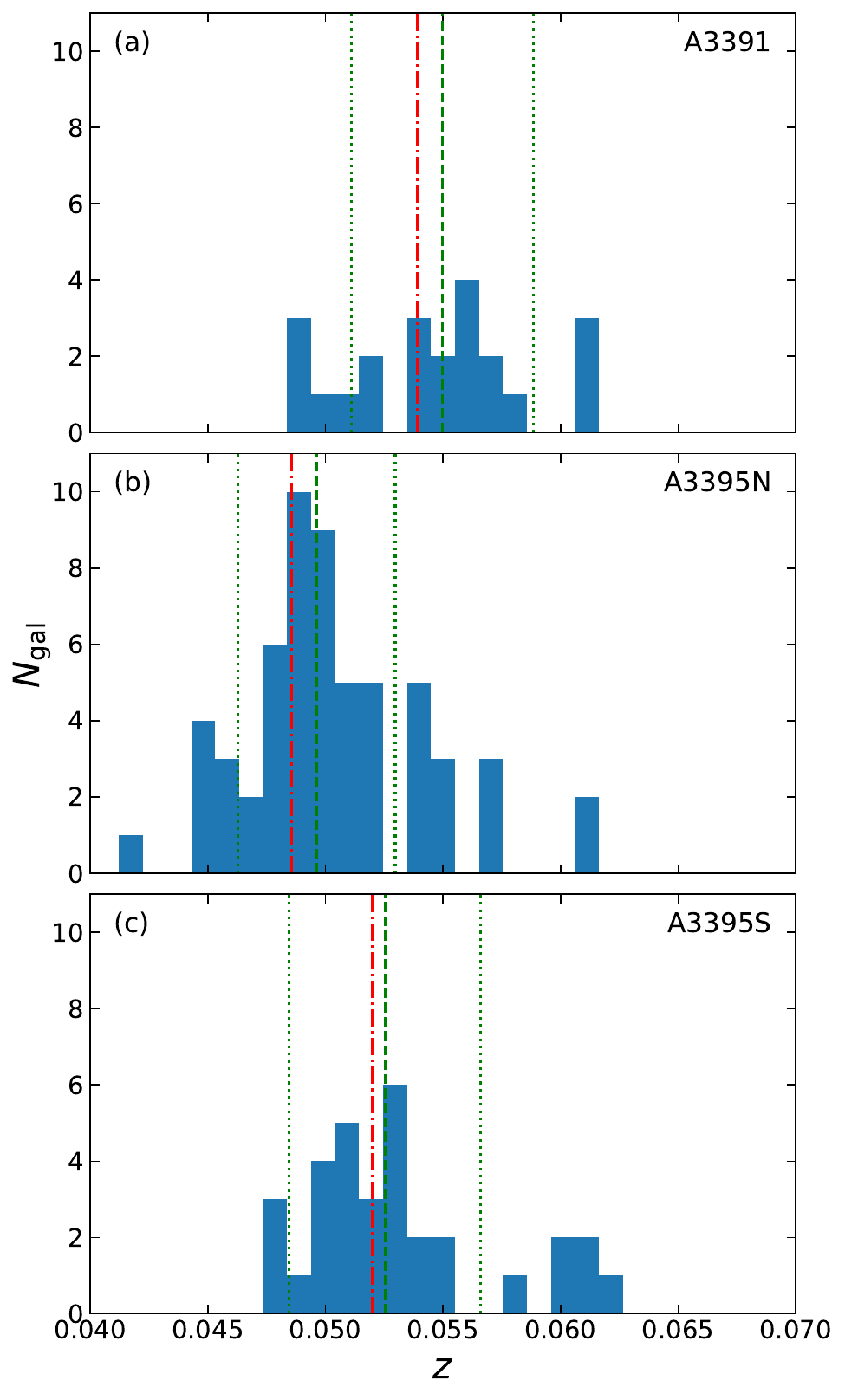}
 \end{center}
 \caption{Redshift distributions of galaxies associated with (a)~A3391, (b)~A3395N, and (c)~A3395S, compiled from NED. The dot-dashed vertical lines indicate the redshifts of the brightest cluster galaxies (BCGs), while the dashed and dotted vertical lines show the median redshift of each cluster and its standard deviation, respectively. {Alt text: Three panels showing galaxy redshift distributions for A3391, A3395N, and A3395S. Each panel displays a distinct redshift peak corresponding to the associated cluster, with differing widths that reflect variations in the galaxy velocity dispersion among the three systems.}
\label{fig:z_hist}}
\end{figure}

Brightest cluster galaxies (BCGs) were identified by visual inspection using optical images from the DESI Legacy Imaging Surveys, overlaid with X-ray surface-brightness contours from eROSITA\footnote{\url{https://erass-cluster-inspector.com/}}\citep{Kluge24}. For each cluster component, the brightest galaxy located near the X-ray peak was selected as the BCG candidate. The BCG redshifts were used as representative values for the galaxy component and compared with the ICM redshift derived from XRISM/Resolve spectroscopy. The resulting galaxy redshift distributions for A3395S, A3395N, and A3391 are shown in figure~\ref{fig:z_hist}.
The corresponding median redshifts, standard errors, and BCG redshifts for each component are summarized in table~\ref{tab:specz}.

\begin{table}[htb]
 \tbl{Optical spectroscopic redshifts}{
    \begin{tabular}{llllll}\hline\hline
Component & $N_{\rm gal}$$^{\mathrm{a}}$ & $\langle z \rangle$$^{\mathrm{b}}$  & $\sigma_{z}$$^{\mathrm{b}}$ & $z_{\rm BCG}$ \\\hline
A3391  & 22 & 0.0550 & 0.0008 & 0.0539 \\ 
A3395N & 56 & 0.0496 & 0.0004 & 0.04856 \\
A3395S & 32 & 0.0525 & 0.0007 & 0.05198 \\\hline
    \end{tabular}}\label{tab:specz}
\begin{tabnote}
$^{\mathrm{a}}$ Number of galaxies used to derive the cluster redshift.  
$^{\mathrm{b}}$ Median spectroscopic redshift of cluster member galaxies and its standard error.
\end{tabnote}
\end{table}

The optical redshift information derived here is used in section~\ref{sec:analysis} to quantify the line-of-sight bulk velocity of the ICM relative to the galaxy component and to assess the dynamical state of the A3395S--A3395N system.

\section{Analysis and results}\label{sec:analysis}
\subsection{Spectral analysis}\label{subsec:spec_analysis}
\begin{figure*}[htb]
 \begin{center}
\includegraphics[width=0.48\linewidth,angle=0]{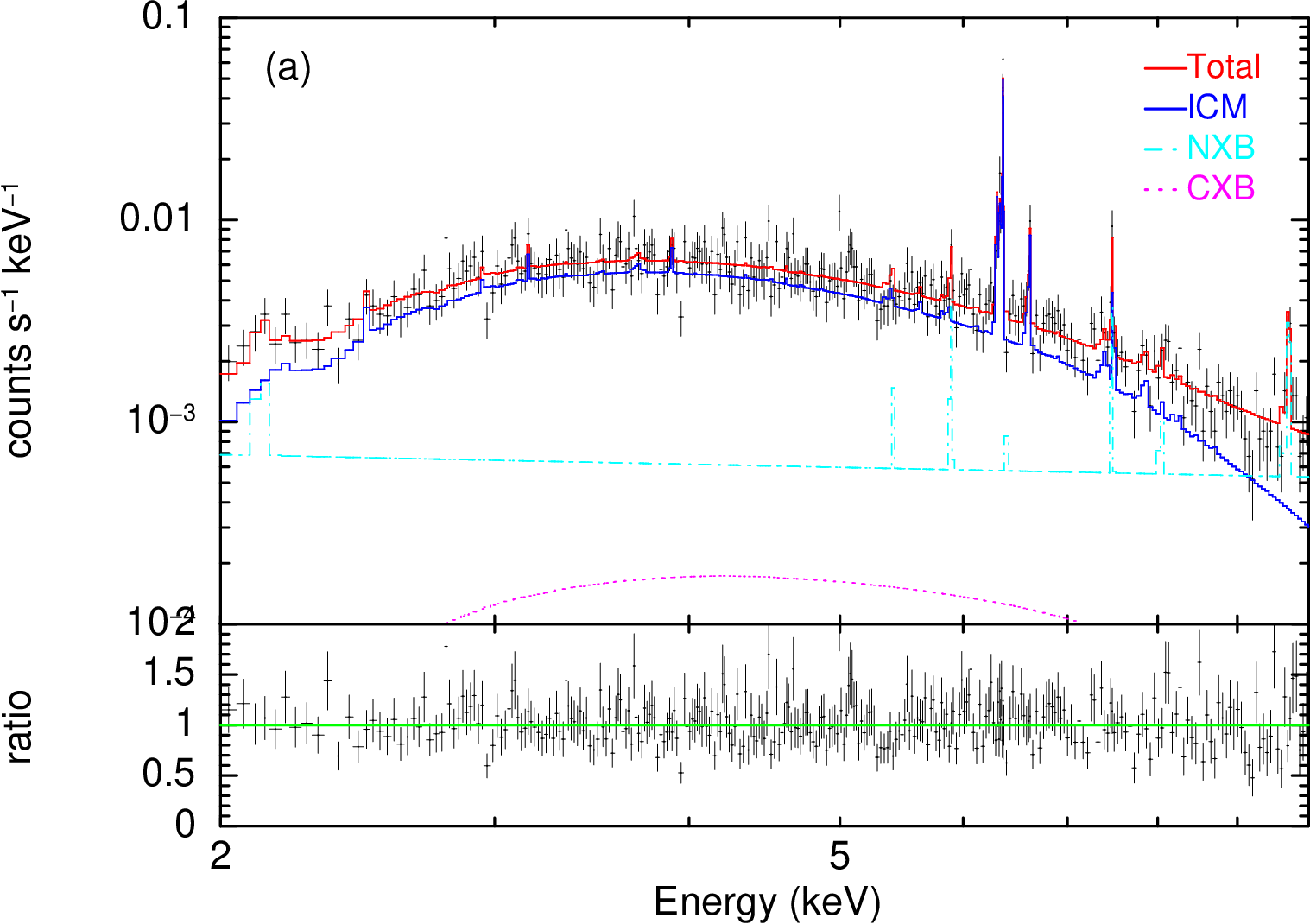}
\includegraphics[width=0.48\linewidth,angle=0]{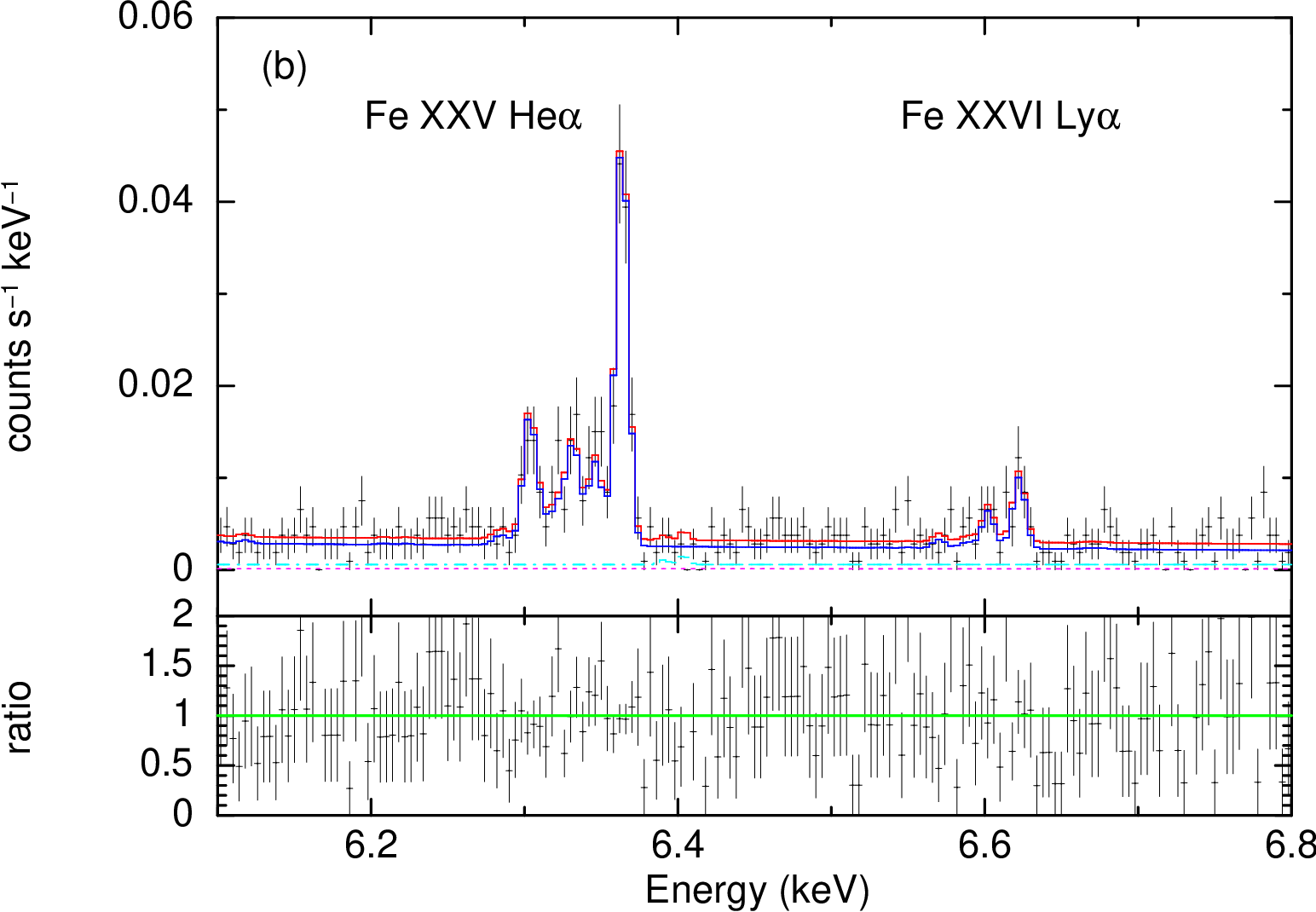}
 \end{center}
 \caption{(a) Resolve spectrum of A3395S in the 2--10~keV band fitted with the \texttt{bapec} model. (b) Zoom-in of the Fe~\textsc{xxv} He$\alpha$ and Fe~\textsc{xxvi} Ly$\alpha$ line complex. The black crosses represent the data points. In panel (a), the spectrum is grouped to achieve a statistical significance of 5$\sigma$ per bin for display purposes. In panel (b), the spectrum is grouped to a fixed bin width, with one bin consisting of eight original channels (4~eV per channel). The red solid line shows the total best-fit model, while the blue solid, cyan dashed, and magenta dotted curves indicate the ICM emission, the non--X-ray background (NXB), and the cosmic X-ray background (CXB) components, respectively. {Alt text: Two-panel X-ray spectra of A3395S. Panel (a) presents the overall spectral shape in the hard X-ray band, while panel (b) focuses on the iron emission line complex, illustrating the detailed line structure used to constrain the kinematic properties of the intracluster medium.}}\label{fig:spec}
\end{figure*}

The Resolve spectrum of A3395S clearly reveals narrow Fe~\textsc{xxv} He$\alpha$ and Fe~\textsc{xxvi} Ly$\alpha$ emission lines (Figure~\ref{fig:spec}). Here we quote the rest-frame energies of the strongest transitions, namely Fe~\textsc{xxv} He$\alpha$ $w$ and Fe~\textsc{xxvi} Ly$\alpha_1$, as 6.7004 and 6.9732~keV, respectively, based on AtomDB version~3.1.3 \citep{Foster12}. As summarized in table~\ref{tab:observation_log}, 224 and 28 counts are detected in the Fe~\textsc{xxv} He$\alpha$ and Fe~\textsc{xxvi} Ly$\alpha$ lines, respectively.

Spectral analysis was performed using the high-resolution Resolve data. Spectra were extracted from the entire Resolve field of view, excluding pixel~27. The spectral fitting was carried out in the 2--10~keV energy band. This band includes the Fe~\textsc{xxv} He$\alpha$ and Fe~\textsc{xxvi} Ly$\alpha$ emission lines, which provide the primary constraints on the temperature, redshift, and velocity broadening of the ICM. The wide energy range was adopted to simultaneously constrain the thermal continuum and the Fe line complex. The impact of restricting the fit to the Fe line band is discussed in section~\ref{subsec:systematic}.

The Resolve spectra were corrected for the barycentric motion of the spacecraft relative to the Solar System barycenter. For the XRISM observation of A3395S, this corresponds to a velocity shift of approximately $-3~{\rm km\,s^{-1}}$ at the time of the observation. This correction was applied to ensure accurate measurements of the ICM redshift and line-of-sight bulk velocity, and its magnitude is negligible compared to the statistical uncertainties.

Figure~\ref{fig:spec} presents the Resolve spectrum and the best-fit model, while the resulting parameters are listed in table~\ref{tab:bapec}. The good agreement between the data and the single-temperature model indicates that the measured line widths are not dominated by unresolved multi-temperature structure along the line of sight, and are therefore likely to primarily reflect intrinsic gas motions.  Although the Resolve extraction region does not exactly match those adopted in previous studies, the ICM temperature and metal abundance measured in the central region of A3395S are consistent with earlier XMM-Newton and Chandra results (e.g., \cite{Lakhchaura11}), within the uncertainties. 

\begin{table*}[htb]
 \tbl{BAPEC model fitting results}{
\begin{tabular}{lllllllll}
\hline\hline
Region & $kT$  & $Z$  & Redshift & $v_{\rm bulk}$  & $\sigma_v$ & norm & C-stat/d.o.f. & $\chi^2$/d.o.f.\\
 & (keV) & (solar) & & (km s$^{-1}$) &  (km s$^{-1}$) & ${\rm (10^{-3} cm^{-5})}$ & & \\ \hline
Full FOV & $4.96^{+0.18}_{-0.17}$ & $0.367^{+0.029}_{-0.027}$ & $0.05291^{+0.00006}_{-0.00006}$ & $264^{+22}_{-23}$ & $124^{+20}_{-21}$ & $7.65^{+0.24}_{-0.24}$ & 14634/15994 & 311/301 \\ \hline
\end{tabular}} \label{tab:bapec}
\begin{tabnote}
The uncertainties on the bulk velocity and turbulent velocity include systematic errors (section~\ref{subsec:systematic}), while those on the other parameters represent statistical errors only. As a reference, the chi-squared value was calculated based on the best-fit result using a spectrum rebinned to contain at least 25 counts per bin.
\end{tabnote}
\end{table*}

\subsection{Systematic uncertainties}\label{subsec:systematic}
We investigated several potential sources of systematic uncertainty that could affect the measurements of the line-of-sight bulk and turbulent velocities.

First, we examined the effect of resonant scattering by excluding the He-like Fe resonance line (the $w$ line) from the spectral fitting. The derived bulk and turbulent velocities remained consistent within the $1\sigma$ statistical uncertainties, indicating that resonant scattering does not significantly affect our results.

Second, we tested the dependence on the adopted energy band. In addition to the baseline fit in the 2--10~keV band, we performed spectral fits restricted to the Fe-line band (5.5--7.0~keV). Although the statistical uncertainties increased due to the reduced bandpass, the best-fit velocities were consistent with those obtained from the full energy band.

Third, we assessed the impact of possible multi-temperature structure in the ICM by fitting the spectra with a two-temperature (\texttt{bapec}+\texttt{bapec}) model. The second temperature component was not statistically required, and the derived velocities were unchanged within uncertainties, suggesting that multi-temperature effects do not bias our measurements.

The systematic uncertainties associated with the Resolve energy scale and the instrumental line spread function (LSF) were evaluated using the gain recovery report for OBSID~201020010. The current $1\sigma$ systematic uncertainty of the Resolve energy scale in the 5.4--9.0~keV band is $\pm0.3$~eV, and the observation-specific energy offset inferred from the calibration pixel is 0.12~eV. Adding these terms in quadrature yields a total energy-scale uncertainty of $\sim0.32$~eV, corresponding to a systematic uncertainty in
the line-of-sight bulk velocity of $\sim15~{\rm km\,s^{-1}}$ at 6~keV. In addition, the systematic uncertainty in the instrumental LSF is energy dependent and amounts to approximately 0.13~eV (FWHM) at 6~keV, which translates to an equivalent Gaussian width of $\sigma_E \simeq 0.055$~eV and a systematic uncertainty in the one-dimensional turbulent velocity of $\sigma_{\rm turb,sys} \sim 3~{\rm km\,s^{-1}}$. Both contributions are small compared to the statistical uncertainties of the measured bulk and turbulent velocities and therefore do not affect our conclusions.

Finally, we examined the impact of background modeling. The cosmic X-ray background (CXB) component is significantly smaller than the non--X-ray background (NXB) in the analyzed region. We confirmed that even allowing for a conservative $\pm50\%$ uncertainty in the CXB normalization does not affect the best-fit velocities, owing to the small extraction region and the dominance of the source emission.

\subsection{Gas kinematics and non-thermal pressure}
\label{subsec:gas_kinematics}

Based on the spectral fitting results presented in section~\ref{subsec:spec_analysis}, we investigate the kinematic properties of the ICM in A3395S, focusing on the line-of-sight bulk velocity, the turbulent velocity dispersion, and the associated non-thermal pressure support.

The line-of-sight bulk velocity of the ICM was derived from the best-fit redshift obtained with the \texttt{bapec} model. We applied the barycentric correction to the measured redshift and converted the redshift difference relative to the BCG into a velocity according to $v = c\,(z - z_{\rm BCG})/(1+z_{\rm BCG})$. Using this definition, we measure a bulk velocity of $v_{\rm bulk} = 264^{+22}_{-23}~{\rm km\,s^{-1}}$. 

For comparison, we also computed the bulk velocity relative to the median redshift of A3395S derived from the galaxy redshift distribution. Using $z_{\rm cluster}=0.0525$ as the reference, we obtain $v_{\rm bulk} = 117 \pm 23~{\rm km\,s^{-1}}$. This value is smaller than that obtained when adopting the BCG redshift, reflecting the offset between the BCG and the cluster median. Nevertheless, the bulk velocity remains significantly non-zero, and the conclusion that the ICM exhibits a finite line-of-sight bulk motion is unchanged.

The line-of-sight turbulent velocity dispersion is measured to be $\sigma_v = 124^{+20}_{-21}~{\rm km\,s^{-1}}$. Assuming isotropic turbulence, this corresponds to a three-dimensional turbulent velocity $v_{\rm turb,3D} = \sqrt{3}\,\sigma_v$. The turbulent Mach number is defined as $\mathcal{M}_{\rm turb} = v_{\rm turb,3D}/c_s$, where the adiabatic sound speed is given by $c_s = \sqrt{\gamma kT/(\mu m_p)}$, with $\gamma = 5/3$ and mean molecular weight $\mu = 0.61$. Using the best-fit temperature, we obtain a turbulent Mach number of $\mathcal{M}_{\rm turb} = 0.188^{+0.030}_{-0.031}$, indicating that the turbulent gas motions in the core of A3395S are clearly subsonic. 

The interpretation of the derived turbulent Mach number is not straightforward.
While unresolved bulk motions along the line of sight can contribute to the measured velocity dispersion and bias it high, the measurement is also subject to line-of-sight emissivity weighting, which limits the effective spatial scale probed.
Based on a $\beta$-model description of the gas distribution with parameters $\beta = 0.55$ and $r_c = 114~\mathrm{kpc}$ \citep{Lakhchaura11}, we estimate a representative effective line-of-sight length of $l_{\rm eff} \sim 70~\mathrm{kpc}$ within the Resolve field of view.
Here, following \citet{Zhuravleva12}, $l_{\rm eff}$ is defined as the one-sided half-length enclosing 50\% of the cumulative line-of-sight emission, i.e.,
\[
\frac{\int_{0}^{l_{\rm eff}} \epsilon(l)\,dl}
{\int_{0}^{\infty} \epsilon(l)\,dl}
=0.5,
\]
where $\epsilon \propto n_e^2$.
This corresponds to a full line-of-sight extent of approximately $2l_{\rm eff}\sim140~\mathrm{kpc}$.
Therefore, the observed velocity dispersion primarily reflects emissivity-weighted gas motions on scales of order $\sim100~\mathrm{kpc}$, and larger-scale velocity fluctuations may not be fully captured.
As a result, the measured Mach number cannot be regarded as a strict upper or lower limit on the intrinsic turbulence level, but should instead be interpreted as an emissivity-weighted, scale-dependent estimate of the velocity field.

The non-thermal pressure associated with turbulent motions is estimated as $P_{\rm turb} = \rho v_{\rm turb,3D}^2/3$, where $\rho$ is the gas mass density, and the ratio of turbulent to thermal pressure is given by $P_{\rm turb}/P_{\rm th} = \mu m_p v_{\rm turb,3D}^2/(3kT)$. We find that the turbulent pressure contributes only a small fraction of the total pressure budget, with $P_{\rm turb}/P_{\rm th} = 0.020^{+0.007}_{-0.006}$.

We also estimate the non-thermal pressure by including both the bulk and turbulent velocity components as $P_{\rm nth,tot} = \rho (v_{\rm bulk}^2 + v_{\rm turb,3D}^2)/3$, which yields $P_{\rm nth,tot}/P_{\rm th} = 0.050^{+0.008}_{-0.007}$. Even when both velocity components are taken into account, the non-thermal pressure support remains modest, indicating that gas motions are dynamically subdominant in the core of A3395S and that the ICM within the Resolve field of view is close to hydrostatic equilibrium, despite the ongoing interaction on larger scales.

\section{Discussion}\label{sec:discuss}
\subsection{Subsonic turbulence and comparison with the radio AGN}

The high-resolution XRISM/Resolve spectroscopy reveals that the line-of-sight turbulent velocity in the core region of A3395S is $\sim120~{\rm km\,s^{-1}}$, 
corresponding to subsonic turbulence with $\mathcal{M}_{\rm turb}\sim0.2$. The turbulent Mach number in A3395S appears to be lower than values reported in some merging cluster cores (e.g., Coma, A754, and A3667; \cite{XRISM25b,Omiya26b,Omiya26a}), although this comparison should be interpreted with caution, as the measurements are emission-weighted and sensitive to the effective spatial scale probed along the line of sight.

The measured turbulent velocity is also broadly comparable in magnitude to values reported for the Perseus cluster core \citep{Hitomi16, Hitomi18, XRISM26a}, where a velocity dispersion of $\sim170~{\rm km\,s^{-1}}$ has been measured in the innermost core despite the presence of a central radio AGN.
However, the effective line-of-sight scales probed are not necessarily identical between the two systems.
In the recent XRISM Perseus study, the effective line-of-sight length varies from several tens of kiloparsecs in the innermost core to a few hundred kiloparsecs at larger radii due to emissivity weighting \citep{XRISM26a}.
For A3395S, our estimate based on the $\beta$-model gives $l_{\rm eff}\sim70~\mathrm{kpc}$ in the one-sided definition of \citet{Zhuravleva12}, corresponding to a full line-of-sight extent of $\sim140~\mathrm{kpc}$ (Section~3.3).
Therefore, the A3395S measurement should be interpreted as an emissivity-weighted estimate of gas motions on scales of order $\sim100~\mathrm{kpc}$, rather than as a direct probe of the innermost AGN-dominated scale in Perseus.
The low velocity dispersion thus indicates the absence of strong volume-filling turbulence on these scales, although localized AGN-driven motions on smaller scales could still be diluted within the Resolve spectrum.

A3395S hosts a central FR~I radio galaxy, S1 (PKS~0625--545), associated with the brightest cluster galaxy \citep{Brueggen21}. ASKAP observations show that its jets and lobes extend over several hundred kiloparsecs ($\sim6'$, corresponding to $\sim370$~kpc). Chandra data indicate that the radio structures are located near regions of steep ICM surface-brightness gradients, suggesting interaction with the surrounding gas; however, there is no clear evidence for prominent X-ray cavities or strong shocks, implying relatively gentle AGN feedback at present.

Although turbulence confined to a very small central region cannot be ruled out given the limited spatial resolution of XRISM, the field-of-view-averaged measurement ($3'\times3'$) reflects line-of-sight emissivity-weighted gas motions. Therefore, any highly localized turbulent motions would be strongly diluted in the observed spectrum, and we find no evidence for strong turbulence dominating the core region on these scales.
Localized AGN-driven motions may exist but would be strongly diluted in the emission-weighted spectrum. In this respect, A3395S may represent an earlier evolutionary phase than systems such as M87, where XRISM has resolved enhanced velocity dispersion associated with AGN feedback in the central atmosphere \citep{XRISM26b}.

No diffuse radio halo has been detected in A3395S, although a triangular-shaped diffuse radio feature west of the cluster has been suggested to be a radio relic or re-accelerated plasma related to a radio AGN \citep{Reiprich21,Brueggen21}. Since radio halos are commonly interpreted as signatures of turbulent re-acceleration during mergers, their absence is consistent with the low level of turbulence inferred from XRISM and supports the view that large-scale turbulent motions have not yet fully developed. While this does not exclude past merger activity, it suggests that A3395S and A3395N have not reached a post-merger phase characterized by well-developed turbulence on core-wide scales.

Cosmological simulations further suggest that turbulence in merging clusters can be spatially intermittent rather than volume-filling, which would reduce the velocity dispersion inferred from line broadening when averaged over the bright core \citep{Vazza26}. This effect should be considered when interpreting the low turbulent velocity measured in A3395S.

\subsection{Bulk velocity and merger geometry}
The XRISM/Resolve spectroscopy reveals a significant line-of-sight bulk velocity of the ICM in the core of A3395S, indicating that the hot gas is not at rest with respect to the cluster systemic redshift and that the system is dynamically unrelaxed. This interpretation is supported by optical measurements showing a redshift offset between the galaxy populations of A3395S and A3395N, as well as by the disturbed X-ray morphology and the presence of an X-ray-emitting bridge connecting the two subclusters. Previous studies have also reported an offset between the brightest cluster galaxy and the X-ray centroid in A3395S \citep{Brueggen21}, further supporting a disturbed dynamical state.

The measured bulk velocity represents only the line-of-sight component of the three-dimensional gas motion; the true relative velocity between the interacting subclusters may therefore be larger. A substantial fraction of the merger-driven motion could lie in the plane of the sky. We searched for spatial variations in the bulk velocity by dividing the Resolve field of view into two subregions. Although the best-fit velocities differ, the statistical significance of the difference is low given the limited photon statistics and PSF mixing, and no robust velocity gradient can be established.

On larger scales, the median redshifts of A3395S and A3395N differ by $\Delta z \simeq 0.0029$ (table~2), corresponding to a line-of-sight velocity difference of $\Delta v_{\rm los} \simeq 830~{\rm km\,s^{-1}}$. Together with the X-ray-emitting band between A3395S and A3395N \citep{Reiprich21}, this suggests that the ongoing interaction has a significant line-of-sight component. In such a configuration, the intracluster gas may lag behind the collisionless galaxy and dark matter components, as expected in merging systems with a non-negligible line-of-sight merger geometry.

The combination of a finite bulk velocity and a low turbulent velocity provides an important constraint on the dynamical state of the core. Following \citet{Zhuravleva12}, the ratio between the line-of-sight bulk velocity and the turbulent velocity dispersion, $u_{\rm los}/\sigma_{\rm los}$, serves as an indicator of the dominant spatial scale of gas motions. 
For the full Resolve field of view, we obtain $u_{\rm los}/\sigma_{\rm los} \simeq 2$. 
However, this ratio alone does not uniquely determine the dominant physical scale of gas motions, as its interpretation depends on the underlying velocity structure and, in particular, on its radial behavior \citep{Zhuravleva12}. 
Recent XRISM results for the Coma cluster show that a similar ratio, together with a velocity gradient and weak radial variation of the velocity dispersion, can be explained by merger-driven large-scale motions in an early phase \citep{Zhang26}. 
While our single-pointing observation does not allow us to assess such radial trends, the measured value is broadly consistent with a scenario in which a significant fraction of the kinetic energy resides in coherent, large-scale motions. 
A more quantitative interpretation will require comparisons with tailored numerical simulations that reproduce the specific dynamical state of the A3395 system.

These results suggest that A3395S is in an early-to-intermediate merger stage, where large-scale gas motions are already present but the turbulent cascade has not yet fully developed. Although projection effects and limited photon statistics prevent a detailed reconstruction of the three-dimensional merger geometry, the XRISM measurements demonstrate that bulk motions and turbulence can be decoupled in interacting clusters. 

\subsection{Comparison with numerical simulations}
\label{subsec:comparison_simulation}
The coexistence of low turbulent velocity and significant bulk motion indicates that a substantial fraction of the kinetic energy in A3395S is stored in coherent large-scale flows rather than in fully developed small-scale turbulence within the cluster core.

This interpretation can be discussed in the context of cosmological hydrodynamical simulations of clusters embedded in large-scale filaments. Based on a comparison between SRG/eROSITA observations and Magneticum simulations, \citet{Biffi22a} showed that the Abell~3391--3395 system is embedded in a prominent filamentary structure and that gas motions are strongly influenced by anisotropic accretion and infall along the intercluster axis on scales of several hundred kiloparsecs to megaparsecs. In such a configuration, substantial kinetic energy can be carried by coherent, directional flows and infalling clumps associated with the large-scale structure.

The velocity dispersion measured with XRISM is emission-weighted along the line of sight and is therefore dominated by the densest gas within the central region covered by the Resolve field of view. The observed turbulent velocity thus reflects primarily the dynamical state of the cluster core, rather than the full volume-averaged velocity field on megaparsec scales. Consequently, large-scale bulk flows associated with filamentary accretion may coexist with relatively low velocity dispersion in the dense core if the transfer of kinetic energy from large-scale motions to small-scale turbulence is inefficient or still developing.

In this sense, the low turbulent velocity measured by XRISM does not imply the absence of significant gas motions in the system as a whole. Rather, it suggests that in the current dynamical stage of A3395S, a considerable fraction of the merger- or accretion-driven kinetic energy has not yet been converted into volume-filling turbulence within the core region. A more quantitative comparison, for example through mock XRISM observations derived from simulations and analyzed with the same pipeline as the data, will be required to directly connect the emission-weighted observational constraints with the underlying three-dimensional velocity field. This will be explored in future work.

\section{Summary}\label{sec:summary}
We have investigated the gas motions in the core region of the Abell~3395 South subcluster (A3395S) using high-resolution X-ray spectroscopy with XRISM/Resolve. By analyzing the Fe~XXV He$\alpha$ complex, supplemented by Fe~XXVI Ly$\alpha$, we directly constrained the line-of-sight bulk and turbulent velocities of the intracluster medium (ICM).

We find that the one-dimensional turbulent velocity in the central region of A3395S is $\sim 100~{\rm km\,s^{-1}}$, corresponding to subsonic turbulence with $\mathcal{M}_{\rm turb}\sim0.2$. This value is lower than typically reported for merger cores, although the comparison should be interpreted with some caution because XRISM probes an emissivity-weighted core scale. The low turbulent velocity is notable despite the presence of a central radio AGN.

In contrast, a finite line-of-sight bulk velocity of $\sim 260~{\rm km\,s^{-1}}$ relative to the BCG is detected, indicating that A3395S has not yet reached a dynamically relaxed state and that large-scale coherent gas motions are  still present.

The subsonic level of turbulence is consistent with the non-detection of a radio halo in A3395S, suggesting that turbulent particle reacceleration is currently inefficient in the cluster core.

This study demonstrates the capability of XRISM to directly probe ICM dynamics in merging clusters. Future comparative analyses of A3395N and A3391 will provide further insight into the roles of cluster mergers and AGN feedback in driving gas motions within large-scale structures.


\bibliographystyle{pasj}
\bibliography{ref}

@ARTICLE{Alvarez18,
       author = {{Alvarez}, Gabriella E. and {Randall}, Scott W. and {Bourdin}, Herv{\'e} and {Jones}, Christine and {Holley-Bockelmann}, Kelly},
        title = "{Chandra and XMM-Newton Observations of the Abell 3395/Abell 3391 Intercluster Filament}",
      journal = {\apj},
         year = 2018,
        month = may,
       volume = {858},
       number = {1},
          eid = {44},
        pages = {44},
          doi = {10.3847/1538-4357/aabad0},
archivePrefix = {arXiv},
       eprint = {1802.08688},
 primaryClass = {astro-ph.HE},
       adsurl = {https://ui.adsabs.harvard.edu/abs/2018ApJ...858...44A}
}

@INPROCEEDINGS{Arnaud96,
       author = {{Arnaud}, K.~A.},
        title = "{XSPEC: The First Ten Years}",
    booktitle = {Astronomical Data Analysis Software and Systems V},
         year = 1996,
       editor = {{Jacoby}, George H. and {Barnes}, Jeannette},
       series = {Astronomical Society of the Pacific Conference Series},
       volume = {101},
        month = jan,
        pages = {17},
       adsurl = {https://ui.adsabs.harvard.edu/abs/1996ASPC..101...17A}
}

@ARTICLE{Biffi22a,
       author = {{Biffi}, Veronica and {Dolag}, Klaus and {Reiprich}, Thomas H. and {Veronica}, Angie and {Ramos-Ceja}, Miriam E. and {Bulbul}, Esra and {Ota}, Naomi and {Ghirardini}, Vittorio},
        title = "{The eROSITA view of the Abell 3391/95 field: Case study from the Magneticum cosmological simulation}",
      journal = {\aap},
         year = 2022,
        month = may,
       volume = {661},
          eid = {A17},
        pages = {A17},
          doi = {10.1051/0004-6361/202141107},
archivePrefix = {arXiv},
       eprint = {2106.14542},
 primaryClass = {astro-ph.CO},
       adsurl = {https://ui.adsabs.harvard.edu/abs/2022A&A...661A..17B}
}

@ARTICLE{Biffi22b,
       author = {{Biffi}, Veronica and {ZuHone}, John A. and {Mroczkowski}, Tony and {Bulbul}, Esra and {Forman}, William},
        title = "{The velocity structure of the intracluster medium during a major merger: Simulated microcalorimeter observations}",
      journal = {\aap},
         year = 2022,
        month = jul,
       volume = {663},
          eid = {A76},
        pages = {A76},
          doi = {10.1051/0004-6361/202142764},
archivePrefix = {arXiv},
       eprint = {2201.12370},
 primaryClass = {astro-ph.CO},
       adsurl = {https://ui.adsabs.harvard.edu/abs/2022A&A...663A..76B}
}

@ARTICLE{Boehringer10,
       author = {{B{\"o}hringer}, Hans and {Werner}, Norbert},
        title = "{X-ray spectroscopy of galaxy clusters: studying astrophysical processes in the largest celestial laboratories}",
      journal = {\aapr},
         year = 2010,
        month = feb,
       volume = {18},
       number = {1-2},
        pages = {127-196},
          doi = {10.1007/s00159-009-0023-3},
       adsurl = {https://ui.adsabs.harvard.edu/abs/2010A&ARv..18..127B}
}

@ARTICLE{Brueggen21,
       author = {{Br{\"u}ggen}, M. and {Reiprich}, T.~H. and {Bulbul}, E. and {Koribalski}, B.~S. and {Andernach}, H. and {Rudnick}, L. and {Hoang}, D.~N. and {Wilber}, A.~G. and {Duchesne}, S.~W. and {Veronica}, A. and {Pacaud}, F. and {Hopkins}, A.~M. and {Norris}, R.~P. and {Johnston-Hollitt}, M. and {Brown}, M.~J.~I. and {Bonafede}, A. and {Brunetti}, G. and {Collier}, J.~D. and {Sanders}, J.~S. and {Vardoulaki}, E. and {Venturi}, T. and {Kapinska}, A.~D. and {Marvil}, J.},
        title = "{Radio observations of the merging galaxy cluster system Abell 3391-Abell 3395}",
      journal = {\aap},
         year = 2021,
        month = mar,
       volume = {647},
          eid = {A3},
        pages = {A3},
          doi = {10.1051/0004-6361/202039533},
archivePrefix = {arXiv},
       eprint = {2012.08775},
 primaryClass = {astro-ph.HE},
       adsurl = {https://ui.adsabs.harvard.edu/abs/2021A&A...647A...3B}
}

@ARTICLE{Dolag05,
       author = {{Dolag}, K. and {Vazza}, F. and {Brunetti}, G. and {Tormen}, G.},
        title = "{Turbulent gas motions in galaxy cluster simulations: the role of smoothed particle hydrodynamics viscosity}",
      journal = {\mnras},
         year = 2005,
        month = dec,
       volume = {364},
       number = {3},
        pages = {753-772},
          doi = {10.1111/j.1365-2966.2005.09630.x},
archivePrefix = {arXiv},
       eprint = {astro-ph/0507480},
 primaryClass = {astro-ph},
       adsurl = {https://ui.adsabs.harvard.edu/abs/2005MNRAS.364..753D}
}

@ARTICLE{Forman81,
       author = {{Forman}, W. and {Bechtold}, J. and {Blair}, W. and {Giacconi}, R. and {van Speybroeck}, L. and {Jones}, C.},
        title = "{Einstein imaging observations of clusters with a bimodal mass distribution.}",
      journal = {\apjl},
         year = 1981,
        month = feb,
       volume = {243},
        pages = {L133-L136},
          doi = {10.1086/183459},
       adsurl = {https://ui.adsabs.harvard.edu/abs/1981ApJ...243L.133F}
}

@ARTICLE{Foster12,
   author = {{Foster}, A.~R. and {Ji}, L. and {Smith}, R.~K. and {Brickhouse}, N.~S.
	},
    title = "{Updated Atomic Data and Calculations for X-Ray Spectroscopy}",
  journal = {\apj},
archivePrefix = "arXiv",
   eprint = {1207.0576},
 primaryClass = "astro-ph.HE",
     year = 2012,
    month = sep,
   volume = 756,
      eid = {128},
    pages = {128},
      doi = {10.1088/0004-637X/756/2/128},
   adsurl = {http://ads.nao.ac.jp/abs/2012ApJ...756..128F}
}

@ARTICLE{Gouin21,
       author = {{Gouin}, C. and {Bonnaire}, T. and {Aghanim}, N.},
        title = "{Shape and connectivity of groups and clusters: Effect of the dynamical state and accretion history}",
      journal = {\aap},
         year = 2021,
        month = jul,
       volume = {651},
          eid = {A56},
        pages = {A56},
          doi = {10.1051/0004-6361/202140327},
archivePrefix = {arXiv},
       eprint = {2101.04686},
 primaryClass = {astro-ph.CO},
       adsurl = {https://ui.adsabs.harvard.edu/abs/2021A&A...651A..56G}
}

@ARTICLE{HI4PI16,
       author = {{HI4PI Collaboration} and {Ben Bekhti}, N. and {Fl{\"o}er}, L. and {Keller}, R. and {Kerp}, J. and {Lenz}, D. and {Winkel}, B. and {Bailin}, J. and {Calabretta}, M.~R. and {Dedes}, L. and {Ford}, H.~A. and {Gibson}, B.~K. and {Haud}, U. and {Janowiecki}, S. and {Kalberla}, P.~M.~W. and {Lockman}, F.~J. and {McClure-Griffiths}, N.~M. and {Murphy}, T. and {Nakanishi}, H. and {Pisano}, D.~J. and {Staveley-Smith}, L.},
        title = "{HI4PI: A full-sky H I survey based on EBHIS and GASS}",
      journal = {\aap},
         year = 2016,
        month = oct,
       volume = {594},
          eid = {A116},
        pages = {A116},
          doi = {10.1051/0004-6361/201629178},
archivePrefix = {arXiv},
       eprint = {1610.06175},
 primaryClass = {astro-ph.GA},
       adsurl = {https://ui.adsabs.harvard.edu/abs/2016A&A...594A.116H}
}

@ARTICLE{Hitomi16,
       author = {{Hitomi Collaboration} and {Aharonian}, Felix and {Akamatsu}, Hiroki and {Akimoto}, Fumie and {Allen}, Steven W. and {Anabuki}, Naohisa and {Angelini}, Lorella and {Arnaud}, Keith and {Audard}, Marc and {Awaki}, Hisamitsu and {Axelsson}, Magnus and {Bamba}, Aya and {Bautz}, Marshall and {Blandford}, Roger and {Brenneman}, Laura and {Brown}, Gregory V. and {Bulbul}, Esra and {Cackett}, Edward and {Chernyakova}, Maria and {Chiao}, Meng and {Coppi}, Paolo and {Costantini}, Elisa and {de Plaa}, Jelle and {den Herder}, Jan-Willem and {Done}, Chris and {Dotani}, Tadayasu and {Ebisawa}, Ken and {Eckart}, Megan and {Enoto}, Teruaki and {Ezoe}, Yuichiro and {Fabian}, Andrew C. and {Ferrigno}, Carlo and {Foster}, Adam and {Fujimoto}, Ryuichi and {Fukazawa}, Yasushi and {Furuzawa}, Akihiro and {Galeazzi}, Massimiliano and {Gallo}, Luigi and {Gandhi}, Poshak and {Giustini}, Margherita and {Goldwurm}, Andrea and {Gu}, Liyi and {Guainazzi}, Matteo and {Haba}, Yoshito and {Hagino}, Kouichi and {Hamaguchi}, Kenji and {Harrus}, Ilana and {Hatsukade}, Isamu and {Hayashi}, Katsuhiro and {Hayashi}, Takayuki and {Hayashida}, Kiyoshi and {Hiraga}, Junko and {Hornschemeier}, Ann and {Hoshino}, Akio and {Hughes}, John and {Iizuka}, Ryo and {Inoue}, Hajime and {Inoue}, Yoshiyuki and {Ishibashi}, Kazunori and {Ishida}, Manabu and {Ishikawa}, Kumi and {Ishisaki}, Yoshitaka and {Itoh}, Masayuki and {Iyomoto}, Naoko and {Kaastra}, Jelle and {Kallman}, Timothy and {Kamae}, Tuneyoshi and {Kara}, Erin and {Kataoka}, Jun and {Katsuda}, Satoru and {Katsuta}, Junichiro and {Kawaharada}, Madoka and {Kawai}, Nobuyuki and {Kelley}, Richard and {Khangulyan}, Dmitry and {Kilbourne}, Caroline and {King}, Ashley and {Kitaguchi}, Takao and {Kitamoto}, Shunji and {Kitayama}, Tetsu and {Kohmura}, Takayoshi and {Kokubun}, Motohide and {Koyama}, Shu and {Koyama}, Katsuji and {Kretschmar}, Peter and {Krimm}, Hans and {Kubota}, Aya and {Kunieda}, Hideyo and {Laurent}, Philippe and {Lebrun}, Fran{\c{c}}ois and {Lee}, Shiu-Hang and {Leutenegger}, Maurice and {Limousin}, Olivier and {Loewenstein}, Michael and {Long}, Knox S. and {Lumb}, David and {Madejski}, Grzegorz and {Maeda}, Yoshitomo and {Maier}, Daniel and {Makishima}, Kazuo and {Markevitch}, Maxim and {Matsumoto}, Hironori and {Matsushita}, Kyoko and {McCammon}, Dan and {McNamara}, Brian and {Mehdipour}, Missagh and {Miller}, Eric and {Miller}, Jon and {Mineshige}, Shin and {Mitsuda}, Kazuhisa and {Mitsuishi}, Ikuyuki and {Miyazawa}, Takuya and {Mizuno}, Tsunefumi and {Mori}, Hideyuki and {Mori}, Koji and {Moseley}, Harvey and {Mukai}, Koji and {Murakami}, Hiroshi and {Murakami}, Toshio and {Mushotzky}, Richard and {Nagino}, Ryo and {Nakagawa}, Takao and {Nakajima}, Hiroshi and {Nakamori}, Takeshi and {Nakano}, Toshio and {Nakashima}, Shinya and {Nakazawa}, Kazuhiro and {Nobukawa}, Masayoshi and {Noda}, Hirofumi and {Nomachi}, Masaharu and {O'Dell}, Steve and {Odaka}, Hirokazu and {Ohashi}, Takaya and {Ohno}, Masanori and {Okajima}, Takashi and {Ota}, Naomi and {Ozaki}, Masanobu and {Paerels}, Frits and {Paltani}, Stephane and {Parmar}, Arvind and {Petre}, Robert and {Pinto}, Ciro and {Pohl}, Martin and {Porter}, F. Scott and {Pottschmidt}, Katja and {Ramsey}, Brian and {Reynolds}, Christopher and {Russell}, Helen and {Safi-Harb}, Samar and {Saito}, Shinya and {Sakai}, Kazuhiro and {Sameshima}, Hiroaki and {Sato}, Goro and {Sato}, Kosuke and {Sato}, Rie and {Sawada}, Makoto and {Schartel}, Norbert and {Serlemitsos}, Peter and {Seta}, Hiromi and {Shidatsu}, Megumi and {Simionescu}, Aurora and {Smith}, Randall and {Soong}, Yang and {Stawarz}, Lukasz and {Sugawara}, Yasuharu and {Sugita}, Satoshi and {Szymkowiak}, Andrew and {Tajima}, Hiroyasu and {Takahashi}, Hiromitsu and {Takahashi}, Tadayuki and {Takeda}, Shin'Ichiro and {Takei}, Yoh and {Tamagawa}, Toru and {Tamura}, Keisuke and {Tamura}, Takayuki and {Tanaka}, Takaaki and {Tanaka}, Yasuo and {Tanaka}, Yasuyuki and {Tashiro}, Makoto and {Tawara}, Yuzuru and {Terada}, Yukikatsu and {Terashima}, Yuichi and {Tombesi}, Francesco and {Tomida}, Hiroshi and {Tsuboi}, Yohko and {Tsujimoto}, Masahiro and {Tsunemi}, Hiroshi and {Tsuru}, Takeshi and {Uchida}, Hiroyuki and {Uchiyama}, Hideki and {Uchiyama}, Yasunobu and {Ueda}, Shutaro and {Ueda}, Yoshihiro and {Ueno}, Shiro and {Uno}, Shin'Ichiro and {Urry}, Meg and {Ursino}, Eugenio and {de Vries}, Cor and {Watanabe}, Shin and {Werner}, Norbert},
        title = "{The quiescent intracluster medium in the core of the Perseus cluster}",
      journal = {\nat},
         year = 2016,
        month = jul,
       volume = {535},
       number = {7610},
        pages = {117-121},
          doi = {10.1038/nature18627},
archivePrefix = {arXiv},
       eprint = {1607.04487},
 primaryClass = {astro-ph.GA},
       adsurl = {https://ui.adsabs.harvard.edu/abs/2016Natur.535..117H}
}

@ARTICLE{Hitomi18,
       author = {{Hitomi Collaboration} and {Aharonian}, Felix and {Akamatsu}, Hiroki and {Akimoto}, Fumie and {Allen}, Steven W. and {Angelini}, Lorella and {Audard}, Marc and {Awaki}, Hisamitsu and {Axelsson}, Magnus and {Bamba}, Aya and {Bautz}, Marshall W. and {Blandford}, Roger and {Brenneman}, Laura W. and {Brown}, Gregory V. and {Bulbul}, Esra and {Cackett}, Edward M. and {Canning}, Rebecca E.~A. and {Chernyakova}, Maria and {Chiao}, Meng P. and {Coppi}, Paolo S. and {Costantini}, Elisa and {de Plaa}, Jelle and {de Vries}, Cor P. and {den Herder}, Jan-Willem and {Done}, Chris and {Dotani}, Tadayasu and {Ebisawa}, Ken and {Eckart}, Megan E. and {Enoto}, Teruaki and {Ezoe}, Yuichiro and {Fabian}, Andrew C. and {Ferrigno}, Carlo and {Foster}, Adam R. and {Fujimoto}, Ryuichi and {Fukazawa}, Yasushi and {Furuzawa}, Akihiro and {Galeazzi}, Massimiliano and {Gallo}, Luigi C. and {Gandhi}, Poshak and {Giustini}, Margherita and {Goldwurm}, Andrea and {Gu}, Liyi and {Guainazzi}, Matteo and {Haba}, Yoshito and {Hagino}, Kouichi and {Hamaguchi}, Kenji and {Harrus}, Ilana M. and {Hatsukade}, Isamu and {Hayashi}, Katsuhiro and {Hayashi}, Takayuki and {Hayashi}, Tasuku and {Hayashida}, Kiyoshi and {Hiraga}, Junko S. and {Hornschemeier}, Ann and {Hoshino}, Akio and {Hughes}, John P. and {Ichinohe}, Yuto and {Iizuka}, Ryo and {Inoue}, Hajime and {Inoue}, Shota and {Inoue}, Yoshiyuki and {Ishida}, Manabu and {Ishikawa}, Kumi and {Ishisaki}, Yoshitaka and {Iwai}, Masachika and {Kaastra}, Jelle and {Kallman}, Tim and {Kamae}, Tsuneyoshi and {Kataoka}, Jun and {Katsuda}, Satoru and {Kawai}, Nobuyuki and {Kelley}, Richard L. and {Kilbourne}, Caroline A. and {Kitaguchi}, Takao and {Kitamoto}, Shunji and {Kitayama}, Tetsu and {Kohmura}, Takayoshi and {Kokubun}, Motohide and {Koyama}, Katsuji and {Koyama}, Shu and {Kretschmar}, Peter and {Krimm}, Hans A. and {Kubota}, Aya and {Kunieda}, Hideyo and {Laurent}, Philippe and {Lee}, Shiu-Hang and {Leutenegger}, Maurice A. and {Limousin}, Olivier and {Loewenstein}, Michael and {Long}, Knox S. and {Lumb}, David and {Madejski}, Greg and {Maeda}, Yoshitomo and {Maier}, Daniel and {Makishima}, Kazuo and {Markevitch}, Maxim and {Matsumoto}, Hironori and {Matsushita}, Kyoko and {McCammon}, Dan and {McNamara}, Brian R. and {Mehdipour}, Missagh and {Miller}, Eric D. and {Miller}, Jon M. and {Mineshige}, Shin and {Mitsuda}, Kazuhisa and {Mitsuishi}, Ikuyuki and {Miyazawa}, Takuya and {Mizuno}, Tsunefumi and {Mori}, Hideyuki and {Mori}, Koji and {Mukai}, Koji and {Murakami}, Hiroshi and {Mushotzky}, Richard F. and {Nakagawa}, Takao and {Nakajima}, Hiroshi and {Nakamori}, Takeshi and {Nakashima}, Shinya and {Nakazawa}, Kazuhiro and {Nobukawa}, Kumiko K. and {Nobukawa}, Masayoshi and {Noda}, Hirofumi and {Odaka}, Hirokazu and {Ohashi}, Takaya and {Ohno}, Masanori and {Okajima}, Takashi and {Ota}, Naomi and {Ozaki}, Masanobu and {Paerels}, Frits and {Paltani}, St{\'e}phane and {Petre}, Robert and {Pinto}, Ciro and {Porter}, Frederick S. and {Pottschmidt}, Katja and {Reynolds}, Christopher S. and {Safi-Harb}, Samar and {Saito}, Shinya and {Sakai}, Kazuhiro and {Sasaki}, Toru and {Sato}, Goro and {Sato}, Kosuke and {Sato}, Rie and {Sawada}, Makoto and {Schartel}, Norbert and {Serlemtsos}, Peter J. and {Seta}, Hiromi and {Shidatsu}, Megumi and {Simionescu}, Aurora and {Smith}, Randall K. and {Soong}, Yang and {Stawarz}, {\L}ukasz and {Sugawara}, Yasuharu and {Sugita}, Satoshi and {Szymkowiak}, Andrew and {Tajima}, Hiroyasu and {Takahashi}, Hiromitsu and {Takahashi}, Tadayuki and {Takeda}, Shin'ichiro and {Takei}, Yoh and {Tamagawa}, Toru and {Tamura}, Takayuki and {Tanaka}, Keigo and {Tanaka}, Takaaki and {Tanaka}, Yasuo and {Tanaka}, Yasuyuki T. and {Tashiro}, Makoto S. and {Tawara}, Yuzuru and {Terada}, Yukikatsu and {Terashima}, Yuichi and {Tombesi}, Francesco and {Tomida}, Hiroshi and {Tsuboi}, Yohko and {Tsujimoto}, Masahiro and {Tsunemi}, Hiroshi and {Tsuru}, Takeshi Go and {Uchida}, Hiroyuki and {Uchiyama}, Hideki and {Uchiyama}, Yasunobu and {Ueda}, Shutaro and {Ueda}, Yoshihiro and {Uno}, Shin'ichiro and {Urry}, C. Megan and {Ursino}, Eugenio and {Wang}, Qian H.~S. and {Watanabe}, Shin and {Werner}, Norbert and {Wilkins}, Dan R. and {Williams}, Brian J. and {Yamada}, Shinya and {Yamaguchi}, Hiroya and {Yamaoka}, Kazutaka and {Yamasaki}, Noriko Y. and {Yamauchi}, Makoto and {Yamauchi}, Shigeo and {Yaqoob}, Tahir and {Yatsu}, Yoichi and {Yonetoku}, Daisuke and {Zhuravleva}, Irina and {Zoghbi}, Abderahmen},
        title = "{Atmospheric gas dynamics in the Perseus cluster observed with Hitomi}",
      journal = {\pasj},
         year = 2018,
        month = mar,
       volume = {70},
       number = {2},
          eid = {9},
        pages = {9},
          doi = {10.1093/pasj/psx138},
archivePrefix = {arXiv},
       eprint = {1711.00240},
 primaryClass = {astro-ph.HE},
       adsurl = {https://ui.adsabs.harvard.edu/abs/2018PASJ...70....9H}
}

@ARTICLE{Ishisaki18,
       author = {{Ishisaki}, Y. and {Ezoe}, Y. and {Yamada}, S. and {Ichinohe}, Y. and {Fujimoto}, R. and {Takei}, Y. and {Yasuda}, S. and {Ishida}, M. and {Yamasaki}, N.~Y. and {Maeda}, Y. and {Tsujimoto}, M. and {Iizuka}, R. and {Koyama}, S. and {Noda}, H. and {Tamagawa}, T. and {Sawada}, M. and {Sato}, K. and {Kitamoto}, S. and {Hoshino}, A. and {Brown}, G.~V. and {Eckart}, M.~E. and {Hayashi}, T. and {Kelley}, R.~L. and {Kilbourne}, C.~A. and {Leutenegger}, M.~A. and {Mori}, H. and {Okajima}, T. and {Porter}, F.~S. and {Soong}, Y. and {McCammon}, D. and {Szymkowiak}, A.~E.},
        title = "{Resolve Instrument on X-ray Astronomy Recovery Mission (XARM)}",
      journal = {Journal of Low Temperature Physics},
         year = 2018,
        month = dec,
       volume = {193},
       number = {5-6},
        pages = {991-995},
          doi = {10.1007/s10909-018-1913-4},
       adsurl = {https://ui.adsabs.harvard.edu/abs/2018JLTP..193..991I}
}

@ARTICLE{Kluge24,
       author = {{Kluge}, M. and {Comparat}, J. and {Liu}, A. and {Balzer}, F. and {Bulbul}, E. and {Ider Chitham}, J. and {Ghirardini}, V. and {Garrel}, C. and {Bahar}, Y.~E. and {Artis}, E. and {Bender}, R. and {Clerc}, N. and {Dwelly}, T. and {Fabricius}, M.~H. and {Grandis}, S. and {Hern{\'a}ndez-Lang}, D. and {Hill}, G.~J. and {Joshi}, J. and {Lamer}, G. and {Merloni}, A. and {Nandra}, K. and {Pacaud}, F. and {Predehl}, P. and {Ramos-Ceja}, M.~E. and {Reiprich}, T.~H. and {Salvato}, M. and {Sanders}, J.~S. and {Schrabback}, T. and {Seppi}, R. and {Zelmer}, S. and {Zenteno}, A. and {Zhang}, X.},
        title = "{The SRG/eROSITA All-Sky Survey. Optical identification and properties of galaxy clusters and groups in the western galactic hemisphere}",
      journal = {\aap},
         year = 2024,
        month = aug,
       volume = {688},
          eid = {A210},
        pages = {A210},
          doi = {10.1051/0004-6361/202349031},
archivePrefix = {arXiv},
       eprint = {2402.08453},
 primaryClass = {astro-ph.CO},
       adsurl = {https://ui.adsabs.harvard.edu/abs/2024A&A...688A.210K}
}

@ARTICLE{Lakhchaura11,
       author = {{Lakhchaura}, Kiran and {Singh}, K.~P. and {Saikia}, D.~J. and {Hunstead}, R.~W.},
        title = "{Intracluster Medium of the Merging Cluster A3395}",
      journal = {\apj},
         year = 2011,
        month = dec,
       volume = {743},
       number = {1},
          eid = {78},
        pages = {78},
          doi = {10.1088/0004-637X/743/1/78},
archivePrefix = {arXiv},
       eprint = {1109.5688},
 primaryClass = {astro-ph.CO},
       adsurl = {https://ui.adsabs.harvard.edu/abs/2011ApJ...743...78L}
}

@ARTICLE{Lodders09,
   author = {{Lodders}, K. and {Palme}, H.},
    title = "{Solar System Elemental Abundances in 2009}",
  journal = {Meteoritics and Planetary Science Supplement},
     year = 2009,
    month = sep,
   volume = 72,
    pages = {5154},
   adsurl = {http://ads.nao.ac.jp/abs/2009M%26PSA..72.5154L}
}

@ARTICLE{Markevitch98,
       author = {{Markevitch}, Maxim and {Forman}, William R. and {Sarazin}, Craig L. and {Vikhlinin}, Alexey},
        title = "{The Temperature Structure of 30 Nearby Clusters Observed with ASCA: Similarity of Temperature Profiles}",
      journal = {\apj},
         year = 1998,
        month = aug,
       volume = {503},
       number = {1},
        pages = {77-96},
          doi = {10.1086/305976},
archivePrefix = {arXiv},
       eprint = {astro-ph/9711289},
 primaryClass = {astro-ph},
       adsurl = {https://ui.adsabs.harvard.edu/abs/1998ApJ...503...77M}
}

@ARTICLE{Noda25,
       author = {{Noda}, Hirofumi and {Mori}, Koji and {Tomida}, Hiroshi and {Nakajima}, Hiroshi and {Tanaka}, Takaaki and {Murakami}, Hiroshi and {Uchida}, Hiroyuki and {Suzuki}, Hiromasa and {Kobayashi}, Shogo Benjamin and {Yoneyama}, Tomokage and {Hagino}, Kouichi and {Nobukawa}, Kumiko and {Uchiyama}, Hideki and {Nobukawa}, Masayoshi and {Matsumoto}, Hironori and {Tsuru}, Takeshi Go and {Yamauchi}, Makoto and {Hatsukade}, Isamu and {Odaka}, Hirokazu and {Kohmura}, Takayoshi and {Yamaoka}, Kazutaka and {Yoshida}, Tessei and {Kanemaru}, Yoshiaki and {Hiraga}, Junko and {Dotani}, Tadayasu and {Ozaki}, Masanobu and {Tsunemi}, Hiroshi and {Sato}, Jin and {Takaki}, Toshiyuki and {Terada}, Yuta and {Miyazaki}, Keitaro and {Kusunoki}, Kohei and {Otsuka}, Yoshinori and {Yokosu}, Haruhiko and {Yonemaru}, Wakana and {Ichikawa}, Kazuhiro and {Nakano}, Hanako and {Takemoto}, Reo and {Matsushima}, Tsukasa and {Urase}, Reika and {Kurashima}, Jun and {Fuchi}, Kotomi and {Hayakawa}, Kaito and {Fukuda}, Masahiro and {Kamei}, Takamitsu and {Asahina}, Yoh and {Inoue}, Shun and {Amano}, Yuki and {Aoki}, Yuma and {Ito}, Yamato and {Kamatani}, Tomoya and {Takayama}, Kouta and {Sako}, Takashi and {Yoshimoto}, Marina and {Shima}, Kohei and {Higuchi}, Mayu and {Ninoyu}, Kaito and {Aoki}, Daiki and {Tsunomachi}, Shun and {Hayashida}, Kiyoshi},
        title = "{Soft X-ray Imager of the Xtend system on board XRISM}",
      journal = {\pasj},
         year = 2025,
        month = sep,
       volume = {77},
        pages = {S10-S22},
          doi = {10.1093/pasj/psaf011},
archivePrefix = {arXiv},
       eprint = {2502.08030},
 primaryClass = {astro-ph.IM},
       adsurl = {https://ui.adsabs.harvard.edu/abs/2025PASJ...77S..10N}
}

@ARTICLE{Omiya26a,
       author = {{Omiya}, Yuki and {Ichinohe}, Yuto and {Nakazawa}, Kazuhiro and {Awaki}, Hisamitsu and {Eckert}, Dominique and {Fujita}, Yutaka and {Hatsukade}, Isamu and {Markevitch}, Maxim and {Mernier}, Fran{\c{c}}ois and {Mitsuishi}, Ikuyuki and {Ota}, Naomi and {Simionescu}, Aurora and {Uchida}, Yuusuke and {Ueda}, Shutaro and {Zhuravleva}, Irina and {Zuhone}, John},
        title = "{XRISM Observations of the Prototypical Cold Front in A3667}",
      journal = {\apjl},
         year = 2026,
        month = jan,
       volume = {996},
       number = {1},
          eid = {L15},
        pages = {L15},
          doi = {10.3847/2041-8213/ae2a28},
archivePrefix = {arXiv},
       eprint = {2510.26405},
 primaryClass = {astro-ph.GA},
       adsurl = {https://ui.adsabs.harvard.edu/abs/2026ApJ...996L..15O}
}

@ARTICLE{Omiya26b,
       author = {{Omiya}, Yuki and {Okabe}, Nobuhiro and {Nakazawa}, Kazuhiro and {Ota}, Naomi and {Ichinohe}, Yuto and {Ueda}, Shutaro and {Nguyen-Dang}, Nhan T.},
        title = "{XRISM-Subaru views of Abell 754: Energetic ICM motions revealed by XRISM/Resolve}",
      journal = {\pasj},
         year = 2026,
        month = feb,
          doi = {10.1093/pasj/psag013},
archivePrefix = {arXiv},
       eprint = {2510.16553},
 primaryClass = {astro-ph.GA},
       adsurl = {https://ui.adsabs.harvard.edu/abs/2026PASJ..tmp...19O}
}

@ARTICLE{Ota12,
       author = {{Ota}, Naomi},
        title = "{X-ray spectroscopy of clusters of galaxies}",
      journal = {Research in Astronomy and Astrophysics},
         year = 2012,
        month = aug,
       volume = {12},
       number = {8},
        pages = {973-994},
          doi = {10.1088/1674-4527/12/8/006},
archivePrefix = {arXiv},
       eprint = {1211.0679},
 primaryClass = {astro-ph.CO},
       adsurl = {https://ui.adsabs.harvard.edu/abs/2012RAA....12..973O}
}

@ARTICLE{Ota18,
       author = {{Ota}, Naomi and {Nagai}, Daisuke and {Lau}, Erwin T.},
        title = "{Constraining hydrostatic mass bias of galaxy clusters with high-resolution X-ray spectroscopy}",
      journal = {\pasj},
         year = 2018,
        month = jun,
       volume = {70},
       number = {3},
          eid = {51},
        pages = {51},
          doi = {10.1093/pasj/psy040},
archivePrefix = {arXiv},
       eprint = {1507.02730},
 primaryClass = {astro-ph.CO},
       adsurl = {https://ui.adsabs.harvard.edu/abs/2018PASJ...70...51O}
}

@ARTICLE{Planck13,
       author = {{Planck Collaboration} and {Ade}, P.~A.~R. and {Aghanim}, N. and {Arnaud}, M. and {Ashdown}, M. and {Atrio-Barandela}, F. and {Aumont}, J. and {Baccigalupi}, C. and {Balbi}, A. and {Banday}, A.~J. and {Barreiro}, R.~B. and {Battaner}, J.~G. Bartlett E. and {Benabed}, K. and {Beno{\^\i}t}, A. and {Bernard}, J.-P. and {Bersanelli}, M. and {Bhatia}, R. and {Bikmaev}, I. and {B{\"o}hringer}, H. and {Bonaldi}, A. and {Bond}, J.~R. and {Borrill}, J. and {Bouchet}, F.~R. and {Bourdin}, H. and {Burenin}, R. and {Burigana}, C. and {Cabella}, P. and {Cardoso}, J.-F. and {Castex}, G. and {Catalano}, A. and {Cay{\'o}n}, L. and {Chamballu}, A. and {Chary}, R.-R. and {Chiang}, L.-Y. and {Chon}, G. and {Christensen}, P.~R. and {Clements}, D.~L. and {Colafrancesco}, S. and {Colombo}, L.~P.~L. and {Comis}, B. and {Coulais}, A. and {Crill}, B.~P. and {Cuttaia}, F. and {Danese}, L. and {Davis}, R.~J. and {de Bernardis}, P. and {de Gasperis}, G. and {de Zotti}, G. and {Delabrouille}, J. and {D{\'e}mocl{\`e}s}, J. and {D{\'e}sert}, F.-X. and {Diego}, J.~M. and {Dolag}, K. and {Dole}, H. and {Donzelli}, S. and {Dor{\'e}}, O. and {D{\"o}rl}, U. and {Douspis}, M. and {Dupac}, X. and {Efstathiou}, G. and {En{\ss}lin}, T.~A. and {Eriksen}, H.~K. and {Finelli}, F. and {Flores-Cacho}, I. and {Forni}, O. and {Frailis}, M. and {Franceschi}, E. and {Frommert}, M. and {Ganga}, K. and {G{\'e}nova-Santos}, T. and {Giard}, M. and {Gilfanov}, M. and {Giraud-H{\'e}raud}, Y. and {Gonz{\'a}lez-Nuevo}, J. and {G{\'o}rski}, K.~M. and {Gregorio}, A. and {Gruppuso}, A. and {Hansen}, F.~K. and {Harrison}, D. and {Hempel}, A. and {Henrot-Versill{\'e}}, S. and {Hern{\'a}ndez-Monteagudo}, C. and {Herranz}, D. and {Hildebrandt}, S.~R. and {Hivon}, E. and {Hobson}, M. and {Holmes}, W.~A. and {Hovest}, W. and {Hurier}, G. and {Jaffe}, T.~R. and {Jaffe}, A.~H. and {Jagemann}, T. and {Jones}, W.~C. and {Juvela}, M. and {Khamitov}, I. and {Kisner}, T.~S. and {Kneissl}, R. and {Knoche}, J. and {Knox}, L. and {Kunz}, M. and {Kurki-Suonio}, H. and {Lagache}, G. and {Lamarre}, J.-M. and {Lasenby}, A. and {Lawrence}, C.~R. and {Le Jeune}, M. and {Leonardi}, R. and {Lilje}, P.~B. and {Linden-V{\o}rnle}, M. and {L{\'o}pez-Caniego}, M. and {Lubin}, P.~M. and {Luzzi}, G. and {Mac{\'\i}as-P{\'e}rez}, J.~F. and {Maffei}, B. and {Maino}, D. and {Mandolesi}, N. and {Maris}, M. and {Marleau}, F. and {Marshall}, D.~J. and {Mart{\'\i}nez-Gonz{\'a}lez}, E. and {Masi}, S. and {Massardi}, M. and {Matarrese}, S. and {Matthai}, F. and {Mazzotta}, P. and {Mei}, S. and {Melchiorri}, A. and {Melin}, J.-B. and {Mendes}, L. and {Mennella}, A. and {Mitra}, S. and {Miville-Desch{\`e}nes}, M.-A. and {Moneti}, A. and {Montier}, L. and {Morgante}, G. and {Munshi}, D. and {Murphy}, J.~A. and {Naselsky}, P. and {Nati}, F. and {Natoli}, P. and {N{\o}rgaard-Nielsen}, H.~U. and {Noviello}, F. and {Novikov}, D. and {Novikov}, I. and {Osborne}, S. and {Pajot}, F. and {Paoletti}, D. and {Pasian}, F. and {Patanchon}, G. and {Perdereau}, O. and {Perotto}, L. and {Perrotta}, F. and {Piacentini}, F. and {Piat}, M. and {Pierpaoli}, E. and {Piffaretti}, R. and {Plaszczynski}, S. and {Pointecouteau}, E. and {Polenta}, G. and {Ponthieu}, N. and {Popa}, L. and {Poutanen}, T. and {Pratt}, G.~W. and {Prunet}, S. and {Puget}, J.-L. and {Rachen}, J.~P. and {Rebolo}, R. and {Reinecke}, M. and {Remazeilles}, M. and {Renault}, C. and {Ricciardi}, S. and {Riller}, T. and {Ristorcelli}, I. and {Rocha}, G. and {Roman}, M. and {Rosset}, C. and {Rossetti}, M. and {Rubi{\~n}o-Mart{\'\i}n}, J.~A. and {Rusholme}, B. and {Sandri}, M. and {Savini}, G. and {Schaefer}, B.~M. and {Scott}, D. and {Smoot}, G.~F. and {Starck}, J.-L. and {Sudiwala}, R. and {Sunyaev}, R. and {Sutton}, D. and {Suur-Uski}, A.-S. and {Sygnet}, J.-F. and {Tauber}, J.~A. and {Terenzi}, L. and {Toffolatti}, L. and {Tomasi}, M. and {Tristram}, M. and {Tucci}, M. and {Valenziano}, L. and {Van Tent}, B. and {Vielva}, P. and {Villa}, F.},
        title = "{Planck intermediate results. VIII. Filaments between interacting clusters}",
      journal = {\aap},
         year = 2013,
        month = feb,
       volume = {550},
          eid = {A134},
        pages = {A134},
          doi = {10.1051/0004-6361/201220194},
archivePrefix = {arXiv},
       eprint = {1208.5911},
 primaryClass = {astro-ph.CO},
       adsurl = {https://ui.adsabs.harvard.edu/abs/2013A&A...550A.134P}
}

@ARTICLE{Predehl21,
       author = {{Predehl}, P. and {Andritschke}, R. and {Arefiev}, V. and {Babyshkin}, V. and {Batanov}, O. and {Becker}, W. and {B{\"o}hringer}, H. and {Bogomolov}, A. and {Boller}, T. and {Borm}, K. and {Bornemann}, W. and {Br{\"a}uninger}, H. and {Br{\"u}ggen}, M. and {Brunner}, H. and {Brusa}, M. and {Bulbul}, E. and {Buntov}, M. and {Burwitz}, V. and {Burkert}, W. and {Clerc}, N. and {Churazov}, E. and {Coutinho}, D. and {Dauser}, T. and {Dennerl}, K. and {Doroshenko}, V. and {Eder}, J. and {Emberger}, V. and {Eraerds}, T. and {Finoguenov}, A. and {Freyberg}, M. and {Friedrich}, P. and {Friedrich}, S. and {F{\"u}rmetz}, M. and {Georgakakis}, A. and {Gilfanov}, M. and {Granato}, S. and {Grossberger}, C. and {Gueguen}, A. and {Gureev}, P. and {Haberl}, F. and {H{\"a}lker}, O. and {Hartner}, G. and {Hasinger}, G. and {Huber}, H. and {Ji}, L. and {Kienlin}, A. v. and {Kink}, W. and {Korotkov}, F. and {Kreykenbohm}, I. and {Lamer}, G. and {Lomakin}, I. and {Lapshov}, I. and {Liu}, T. and {Maitra}, C. and {Meidinger}, N. and {Menz}, B. and {Merloni}, A. and {Mernik}, T. and {Mican}, B. and {Mohr}, J. and {M{\"u}ller}, S. and {Nandra}, K. and {Nazarov}, V. and {Pacaud}, F. and {Pavlinsky}, M. and {Perinati}, E. and {Pfeffermann}, E. and {Pietschner}, D. and {Ramos-Ceja}, M.~E. and {Rau}, A. and {Reiffers}, J. and {Reiprich}, T.~H. and {Robrade}, J. and {Salvato}, M. and {Sanders}, J. and {Santangelo}, A. and {Sasaki}, M. and {Scheuerle}, H. and {Schmid}, C. and {Schmitt}, J. and {Schwope}, A. and {Shirshakov}, A. and {Steinmetz}, M. and {Stewart}, I. and {Str{\"u}der}, L. and {Sunyaev}, R. and {Tenzer}, C. and {Tiedemann}, L. and {Tr{\"u}mper}, J. and {Voron}, V. and {Weber}, P. and {Wilms}, J. and {Yaroshenko}, V.},
        title = "{The eROSITA X-ray telescope on SRG}",
      journal = {\aap},
         year = 2021,
        month = mar,
       volume = {647},
          eid = {A1},
        pages = {A1},
          doi = {10.1051/0004-6361/202039313},
archivePrefix = {arXiv},
       eprint = {2010.03477},
 primaryClass = {astro-ph.HE},
       adsurl = {https://ui.adsabs.harvard.edu/abs/2021A&A...647A...1P}
}

@INCOLLECTION{Sanders23,
       author = {{Sanders}, Jeremy S.},
        title = "{Clusters of Galaxies}",
    booktitle = {High-Resolution X-ray Spectroscopy: Instrumentation},
         year = 2023,
       editor = {{Bambi}, Cosimo and {Jiang}, Jiachen},
        pages = {173-207},
          doi = {10.1007/978-981-99-4409-5_8},
       adsurl = {https://ui.adsabs.harvard.edu/abs/2023hxsi.book..173S}
}

@ARTICLE{Reiprich21,
       author = {{Reiprich}, T.~H. and {Veronica}, A. and {Pacaud}, F. and {Ramos-Ceja}, M.~E. and {Ota}, N. and {Sanders}, J. and {Kara}, M. and {Erben}, T. and {Klein}, M. and {Erler}, J. and {Kerp}, J. and {Hoang}, D.~N. and {Br{\"u}ggen}, M. and {Marvil}, J. and {Rudnick}, L. and {Biffi}, V. and {Dolag}, K. and {Aschersleben}, J. and {Basu}, K. and {Brunner}, H. and {Bulbul}, E. and {Dennerl}, K. and {Eckert}, D. and {Freyberg}, M. and {Gatuzz}, E. and {Ghirardini}, V. and {K{\"a}fer}, F. and {Merloni}, A. and {Migkas}, K. and {Nandra}, K. and {Predehl}, P. and {Robrade}, J. and {Salvato}, M. and {Whelan}, B. and {Diaz-Ocampo}, A. and {Hernandez-Lang}, D. and {Zenteno}, A. and {Brown}, M.~J.~I. and {Collier}, J.~D. and {Diego}, J.~M. and {Hopkins}, A.~M. and {Kapinska}, A. and {Koribalski}, B. and {Mroczkowski}, T. and {Norris}, R.~P. and {O'Brien}, A. and {Vardoulaki}, E.},
        title = "{The Abell 3391/95 galaxy cluster system. A 15 Mpc intergalactic medium emission filament, a warm gas bridge, infalling matter clumps, and (re-) accelerated plasma discovered by combining SRG/eROSITA data with ASKAP/EMU and DECam data}",
      journal = {\aap},
         year = 2021,
        month = mar,
       volume = {647},
          eid = {A2},
        pages = {A2},
          doi = {10.1051/0004-6361/202039590},
archivePrefix = {arXiv},
       eprint = {2012.08491},
 primaryClass = {astro-ph.CO},
       adsurl = {https://ui.adsabs.harvard.edu/abs/2021A&A...647A...2R}
}

@ARTICLE{Smith01,
   author = {{Smith}, R.~K. and {Brickhouse}, N.~S. and {Liedahl}, D.~A. and 
	{Raymond}, J.~C.},
    title = "{Collisional Plasma Models with APEC/APED: Emission-Line Diagnostics of Hydrogen-like and Helium-like Ions}",
  journal = {\apjl},
   eprint = {astro-ph/0106478},
     year = 2001,
    month = aug,
   volume = 556,
    pages = {L91-L95},
      doi = {10.1086/322992},
   adsurl = {http://ads.nao.ac.jp/abs/2001ApJ...556L..91S}
}

@ARTICLE{Sugawara17,
       author = {{Sugawara}, Yuuki and {Takizawa}, Motokazu and {Itahana}, Madoka and {Akamatsu}, Hiroki and {Fujita}, Yutaka and {Ohashi}, Takaya and {Ishisaki}, Yoshitaka},
        title = "{Suzaku observations of the outskirts of the galaxy cluster Abell 3395, including a filament toward Abell 3391}",
      journal = {\pasj},
         year = 2017,
        month = dec,
       volume = {69},
       number = {6},
          eid = {93},
        pages = {93},
          doi = {10.1093/pasj/psx104},
archivePrefix = {arXiv},
       eprint = {1708.09074},
 primaryClass = {astro-ph.CO},
       adsurl = {https://ui.adsabs.harvard.edu/abs/2017PASJ...69...93S}
}

@ARTICLE{Tashiro25,
       author = {{Tashiro}, Makoto and {Kelley}, Richard and {Watanabe}, Shin and {Maejima}, Hironori and {Reichenthal}, Lillian and {Toda}, Kenichi and {Hartz}, Leslie and {Santovincenzo}, Andrea and {Matsushita}, Kyoko and {Yamaguchi}, Hiroya and {Petre}, Robert and {Williams}, Brian and {Guainazzi}, Matteo and {Costantini}, Elisa and {Takei}, Yoh and {Ishisaki}, Yoshitaka and {Fujimoto}, Ryuichi and {Henegar-Leon}, Joy and {Sneiderman}, Gary and {Tomida}, Hiroshi and {Mori}, Koji and {Nakajima}, Hiroshi and {Terada}, Yukikatsu and {Holland}, Matthew and {Loewenstein}, Michael and {Miller}, Eric and {Sawada}, Makoto and {Kallman}, Timothy and {Kaastra}, Jelle and {Done}, Chris and {Enoto}, Teruaki and {Bamba}, Aya and {Corrales}, Lia and {Ueda}, Yoshihiro and {Kara}, Erin and {Zhuravleva}, Irina and {Fujita}, Yutaka and {Arai}, Yoshitaka and {Audard}, Marc and {Awaki}, Hisamitsu and {Ballhausen}, Ralf and {Baluta}, Chris and {Bando}, Nobutaka and {Behar}, Ehud and {Bialas}, Thomas and {Boissay-Malaquin}, Rozenn and {Brenneman}, Laura and {Brown}, Gregory V. and {Chiao}, Meng and {Cumbee}, Renata and {de Vries}, Cor and {den Herder}, Jan-Willem and {D{\'\i}az Trigo}, Mar{\'\i}a and {DiPirro}, Michael and {Dotani}, Tadayasu and {Carrero}, Jacobo Ebrero and {Ebisawa}, Ken and {Eckart}, Megan and {Eckert}, Dominique and {Eguchi}, Satoshi and {Ezoe}, Yuichiro and {Ferrigno}, Carlo and {Foster}, Adam and {Fukazawa}, Yasushi and {Fukushima}, Kotaro and {Furuzawa}, Akihiro and {Gallo}, Luigi C. and {Garcia Martinez}, Javier and {Gorter}, Nathalie and {Grim}, Martin and {Gu}, Liyi and {Hagino}, Kouichi and {Hamaguchi}, Kenji and {Hatsukade}, Isamu and {Hayashi}, Katsuhiro and {Hayashi}, Takayuki and {Hell}, Natalie and {Hodges-Kluck}, Edmund and {Horiuchi}, Takafumi and {Hornschemeier}, Ann and {Hoshino}, Akio and {Ichinohe}, Yuto and {Ikuta}, Chisato and {Iizuka}, Ryo and {Ishi}, Daiki and {Ishida}, Manabu and {Ishihama}, Naoki and {Ishikawa}, Kumi and {Ishimura}, Kosei and {Jaffe}, Tess and {Katsuda}, Satoru and {Kanemaru}, Yoshiaki and {Kenyon}, Steven and {Kilbourne}, Caroline and {Kimball}, Mark and {Kitamoto}, Shunji and {Kobayashi}, Shogo and {Kohmura}, Takayoshi and {Kubota}, Aya and {Leutenegger}, Maurice A. and {Maeda}, Yoshitomo and {Markevitch}, Maxim and {Matsumoto}, Hironori and {Matsuzaki}, Keiichi and {McCammon}, Dan and {McLaughlin}, Brian and {McNamara}, Brian and {Mernier}, Fran{\c{c}}ois and {Miko}, Joseph and {Miller}, Jon M. and {Minesugi}, Kenji and {Mitani}, Shinji and {Mitsuishi}, Ikuyuki and {Mizumoto}, Misaki and {Mizuno}, Tsunefumi and {Mukai}, Koji and {Murakami}, Hiroshi and {Mushotzky}, Richard and {Nakazawa}, Kazuhiro and {Natsukari}, Chikara and {Ness}, Jan-Uwe and {Nigo}, Kenichiro and {Nishiyama}, Mari and {Nobukawa}, Kumiko and {Nobukawa}, Masayoshi and {Noda}, Hirofumi and {Odaka}, Hirokazu and {Ogawa}, Mina and {Ogawa}, Shoji and {Ogorzalek}, Anna and {Okajima}, Takashi and {Okamoto}, Atsushi and {Ota}, Naomi and {Ozaki}, Masanobu and {Paltani}, Stephane and {Plucinsky}, Paul and {Porter}, F. Scott and {Pottschmidt}, Katja and {Quero}, Jose Antonio and {Sasaki}, Takahiro and {Sato}, Kosuke and {Sato}, Rie and {Sato}, Toshiki and {Sato}, Yoichi and {Seta}, Hiromi and {Shida}, Maki and {Shidatsu}, Megumi and {Shigeto}, Shuhei and {Shipman}, Russel and {Shinozaki}, Keisuke and {Shirron}, Peter and {Simionescu}, Aurora and {Smith}, Randall K. and {Soong}, Yang and {Suzuki}, Hiromasa and {Szymkowiak}, Andrew and {Takahashi}, Hiromitsu and {Takeo}, Mai and {Tamagawa}, Toru and {Tamura}, Keisuke and {Tanaka}, Takaaki and {Tanimoto}, Atsushi and {Terashima}, Yuichi and {Tsuboi}, Yohko and {Tsujimoto}, Masahiro and {Tsunemi}, Hiroshi and {Tsuru}, Takeshi Go and {Uchida}, Hiroyuki and {Uchida}, Nagomi and {Uchida}, Yuusuke and {Uchiyama}, Hideki and {Uno}, Shinichiro and {Vink}, Jacco and {Witthoeft}, Michael and {Wolfs}, Rob and {Yamada}, Satoshi and {Yamada}, Shinya and {Yamaoka}, Kazutaka and {Yamasaki}, Noriko and {Yamauchi}, Makoto and {Yamauchi}, Shigeo and {Yanagase}, Keiichi and {Yaqoob}, Tahir and {Yasuda}, Susumu and {Yoneyama}, Tomokage and {Yoshida}, Tessei and {Yukita}, Mihoko},
        title = "{X-Ray Imaging and Spectroscopy Mission}",
      journal = {\pasj},
         year = 2025,
        month = sep,
       volume = {77},
        pages = {S1-S9},
          doi = {10.1093/pasj/psaf023},
       adsurl = {https://ui.adsabs.harvard.edu/abs/2025PASJ...77S...1T}
}

@ARTICLE{Tittley01,
       author = {{Tittley}, Eric R. and {Henriksen}, Mark},
        title = "{A Filament between Galaxy Clusters A3391 and A3395}",
      journal = {\apj},
         year = 2001,
        month = dec,
       volume = {563},
       number = {2},
        pages = {673-686},
          doi = {10.1086/323955},
       adsurl = {https://ui.adsabs.harvard.edu/abs/2001ApJ...563..673T}
}

@ARTICLE{Uchida25,
       author = {{Uchida}, Hiroyuki and {Mori}, Koji and {Tomida}, Hiroshi and {Nakajima}, Hiroshi and {Noda}, Hirofumi and {Tanaka}, Takaaki and {Murakami}, Hiroshi and {Suzuki}, Hiromasa and {Kobayashi}, Shogo Benjamin and {Yoneyama}, Tomokage and {Hagino}, Kouichi and {Nobukawa}, Kumiko Kawabata and {Uchiyama}, Hideki and {Nobukawa}, Masayoshi and {Matsumoto}, Hironori and {Tsuru}, Takeshi Go and {Yamauchi}, Makoto and {Hatsukade}, Isamu and {Odaka}, Hirokazu and {Kohmura}, Takayoshi and {Yamaoka}, Kazutaka and {Yoshida}, Tessei and {Kanemaru}, Yoshiaki and {Ishi}, Daiki and {Dotani}, Tadayasu and {Ozaki}, Masanobu and {Tsunemi}, Hiroshi and {Miyazaki}, Keitaro and {Kusunoki}, Kohei and {Otsuka}, Yoshinori and {Yokosu}, Haruhiko and {Yonemaru}, Wakana and {Ichikawa}, Kazuhiro and {Nakano}, Hanako and {Takemoto}, Reo and {Matsushima}, Tsukasa and {Urase}, Reika and {Kurashima}, Jun and {Fuchi}, Kotomi and {Hayakawa}, Kaito and {Fukuda}, Masahiro and {Inoue}, Shun and {Aoki}, Yuma and {Takayama}, Kouta and {Sako}, Takashi and {Yoshimoto}, Marina and {Shima}, Kohei and {Higuchi}, Mayu and {Ninoyu}, Kaito and {Aoki}, Daiki and {Tsunomachi}, Shun and {Okajima}, Takashi and {Ishida}, Manabu and {Maeda}, Yoshitomo and {Hayashi}, Takayuki and {Tamura}, Keisuke and {Boissay-Malaquin}, Rozenn and {Sato}, Toshiki and {Takeo}, Mai and {Miyamoto}, Asca and {Matsumoto}, Gakuto and {Eckart}, Megan E. and {Hell}, Natalie and {Leutenegger}, Maurice A. and {Hayashida}, Kiyoshi},
        title = "{In-orbit performance of the soft X-ray imaging telescope Xtend aboard XRISM}",
      journal = {\pasj},
         year = 2025,
        month = sep,
       volume = {77},
        pages = {S23-S38},
          doi = {10.1093/pasj/psaf030},
archivePrefix = {arXiv},
       eprint = {2503.20180},
 primaryClass = {astro-ph.IM},
       adsurl = {https://ui.adsabs.harvard.edu/abs/2025PASJ...77S..23U}
}

@ARTICLE{Vazza26,
       author = {{Vazza}, F. and {Brunetti}, G.},
        title = "{The interpretation of XRISM X-ray measurements of turbulence in the intracluster medium: A comparison with cosmological simulations}",
      journal = {\aap},
         year = 2026,
        month = jan,
       volume = {705},
          eid = {A129},
        pages = {A129},
          doi = {10.1051/0004-6361/202556070},
archivePrefix = {arXiv},
       eprint = {2507.04727},
 primaryClass = {astro-ph.CO},
       adsurl = {https://ui.adsabs.harvard.edu/abs/2026A&A...705A.129V}
}

@ARTICLE{Veronica24,
       author = {{Veronica}, Angie and {Reiprich}, Thomas H. and {Pacaud}, Florian and {Ota}, Naomi and {Aschersleben}, Jann and {Biffi}, Veronica and {Bulbul}, Esra and {Clerc}, Nicolas and {Dolag}, Klaus and {Erben}, Thomas and {Gatuzz}, Efrain and {Ghirardini}, Vittorio and {Kerp}, J{\"u}rgen and {Klein}, Matthias and {Liu}, Ang and {Liu}, Teng and {Migkas}, Konstantinos and {Ramos-Ceja}, Miriam E. and {Sanders}, Jeremy and {Spinelli}, Claudia},
        title = "{The eROSITA view of the Abell 3391/95 field. Cluster outskirts and filaments}",
      journal = {\aap},
         year = 2024,
        month = jan,
       volume = {681},
          eid = {A108},
        pages = {A108},
          doi = {10.1051/0004-6361/202347037},
archivePrefix = {arXiv},
       eprint = {2311.07488},
 primaryClass = {astro-ph.CO},
       adsurl = {https://ui.adsabs.harvard.edu/abs/2024A&A...681A.108V}
}

@ARTICLE{2022A&A...661A..46V,
       author = {{Veronica}, Angie and {Su}, Yuanyuan and {Biffi}, Veronica and {Reiprich}, Thomas H. and {Pacaud}, Florian and {Nulsen}, Paul E.~J. and {Kraft}, Ralph P. and {Sanders}, Jeremy S. and {Bogdan}, Akos and {Kara}, Melih and et al.},
        title = "The {eROSITA} view of the {A}bell 3391/95 field: The Northern Clump. The largest infalling structure in the longest known gas filament observed with {eROSITA}, {XMM-N}ewton, and {C}handra",
      journal = {\aap},
         year = 2022,
        month = may,
       volume = {661},
          eid = {A46},
        pages = {A46},
          doi = {10.1051/0004-6361/202141415},
archivePrefix = {arXiv},
       eprint = {2106.14543},
 primaryClass = {astro-ph.CO},
       adsurl = {https://ui.adsabs.harvard.edu/abs/2022A&A...661A..46V}
}

@ARTICLE{XRISM25b,
       author = {{XRISM Collaboration} and {Audard}, Marc and {Awaki}, Hisamitsu and {Ballhausen}, Ralf and {Bamba}, Aya and {Behar}, Ehud and {Boissay-Malaquin}, Rozenn and {Brenneman}, Laura and {Brown}, Gregory V. and {Corrales}, Lia and {Costantini}, Elisa and {Cumbee}, Renata and {Diaz Trigo}, Maria and {Done}, Chris and {Dotani}, Tadayasu and {Ebisawa}, Ken and {Eckart}, Megan E. and {Eckert}, Dominique and {Eguchi}, Satoshi and {Enoto}, Teruaki and {Ezoe}, Yuichiro and {Foster}, Adam and {Fujimoto}, Ryuichi and {Fujita}, Yutaka and {Fukazawa}, Yasushi and {Fukushima}, Kotaro and {Furuzawa}, Akihiro and {Gallo}, Luigi and {Garc{\'\i}a}, Javier A. and {Gu}, Liyi and {Guainazzi}, Matteo and {Hagino}, Kouichi and {Hamaguchi}, Kenji and {Hatsukade}, Isamu and {Hayashi}, Katsuhiro and {Hayashi}, Takayuki and {Hell}, Natalie and {Hodges-Kluck}, Edmund and {Hornschemeier}, Ann and {Ichinohe}, Yuto and {Ishi}, Daiki and {Ishida}, Manabu and {Ishikawa}, Kumi and {Ishisaki}, Yoshitaka and {Kaastra}, Jelle and {Kallman}, Timothy and {Kara}, Erin and {Katsuda}, Satoru and {Kanemaru}, Yoshiaki and {Kelley}, Richard and {Kilbourne}, Caroline and {Kitamoto}, Shunji and {Kobayashi}, Shogo and {Kohmura}, Takayoshi and {Kubota}, Aya and {Leutenegger}, Maurice and {Loewenstein}, Michael and {Maeda}, Yoshitomo and {Markevitch}, Maxim and {Matsumoto}, Hironori and {Matsushita}, Kyoko and {McCammon}, Dan and {McNamara}, Brian and {Mernier}, Fran{\c{c}}ois and {Miller}, Eric D. and {Miller}, Jon M. and {Mitsuishi}, Ikuyuki and {Mizumoto}, Misaki and {Mizuno}, Tsunefumi and {Mori}, Koji and {Mukai}, Koji and {Murakami}, Hiroshi and {Mushotzky}, Richard and {Nakajima}, Hiroshi and {Nakazawa}, Kazuhiro and {Ness}, Jan-Uwe and {Nobukawa}, Kumiko and {Nobukawa}, Masayoshi and {Noda}, Hirofumi and {Odaka}, Hirokazu and {Ogawa}, Shoji and {Ogorza{\l}ek}, Anna and {Okajima}, Takashi and {Ota}, Naomi and {Paltani}, St{\'e}phane and {Petre}, Robert and {Plucinsky}, Paul and {Porter}, Frederick S. and {Pottschmidt}, Katja and {Sato}, Kosuke and {Sato}, Toshiki and {Sawada}, Makoto and {Seta}, Hiromi and {Shidatsu}, Megumi and {Simionescu}, Aurora and {Smith}, Randall and {Suzuki}, Hiromasa and {Szymkowiak}, Andrew and {Takahashi}, Hiromitsu and {Takeo}, Mai and {Tamagawa}, Toru and {Tamura}, Keisuke and {Tanaka}, Takaaki and {Tanimoto}, Atsushi and {Tashiro}, Makoto and {Terada}, Yukikatsu and {Terashima}, Yuichi and {Tsuboi}, Yohko and {Tsujimoto}, Masahiro and {Tsunemi}, Hiroshi and {Tsuru}, Takeshi and {T{\"u}mer}, Ay{\c{s}}eg{\"u}l and {Uchida}, Hiroyuki and {Uchida}, Nagomi and {Uchida}, Yuusuke and {Uchiyama}, Hideki and {Ueda}, Shutaro and {Ueda}, Yoshihiro and {Uno}, Shinichiro and {Vink}, Jacco and {Watanabe}, Shin and {Williams}, Brian J. and {Yamada}, Satoshi and {Yamada}, Shinya and {Yamaguchi}, Hiroya and {Yamaoka}, Kazutaka and {Yamasaki}, Noriko and {Yamauchi}, Makoto and {Yamauchi}, Shigeo and {Yaqoob}, Tahir and {Yoneyama}, Tomokage and {Yoshida}, Tessei and {Yukita}, Mihoko and {Zhuravleva}, Irina and {Fabian}, Andrew and {Nelson}, Dylan and {Okabe}, Nobuhiro and {Pillepich}, Annalisa and {Potter}, Cicely and {Regamey}, Manon and {Sakai}, Kosei and {Shishido}, Mona and {Truong}, Nhut and {Wik}, Daniel R. and {Zuhone}, John},
        title = "{XRISM Forecast for the Coma Cluster: Stormy, with a Steep Power Spectrum}",
      journal = {\apjl},
         year = 2025,
        month = may,
       volume = {985},
       number = {1},
          eid = {L20},
        pages = {L20},
          doi = {10.3847/2041-8213/add2f6},
archivePrefix = {arXiv},
       eprint = {2504.20928},
 primaryClass = {astro-ph.HE},
       adsurl = {https://ui.adsabs.harvard.edu/abs/2025ApJ...985L..20X}
}

\begin{ack}
We thank the XRISM project and the mission operations team for the operation of the satellite and their support of the observations. We thank the anonymous referee for careful reading of the manuscript and constructive comments. This work was supported by JSPS KAKENHI Grant Number 25K01026, 23H00121 (NO). This work was also supported by the Humboldt Japan-Germany Fellowship for Joint Research (NO, AV). NO acknowledges partial support by the Organization for the Promotion of Gender Equality at Nara Women's University. AY acknowledges support by JST SPRING, Grant Number JPMJSP2115. MB acknowledges funding by the Deutsche Forschungsgemeinschaft (DFG) under Germany's Excellence Strategy -- EXC 2121 ``Quantum Universe" --  390833306 and the DFG Research Group "Relativistic Jets". This research has made use of the NASA/IPAC Extragalactic Database (NED), which is operated by the Jet Propulsion Laboratory, California Institute of Technology, under contract with the National Aeronautics and Space Administration. 

This work is based on data from eROSITA, the soft X-ray instrument aboard SRG, a joint Russian-German science mission supported by the Russian Space Agency (Roskosmos), in the interests of the Russian Academy of Sciences represented by its Space Research Institute (IKI), and the Deutsches Zentrum f\"{u}r Luft- und Raumfahrt (DLR). The SRG spacecraft was built by Lavochkin Association (NPOL) and its subcontractors, and is operated by NPOL with support from the Max Planck Institute for Extraterrestrial Physics (MPE).

The development and construction of the eROSITA X-ray instrument was led by MPE, with contributions from the Dr. Karl Remeis Observatory Bamberg \& ECAP (FAU Erlangen-Nuernberg), the University of Hamburg Observatory, the Leibniz Institute for Astrophysics Potsdam (AIP), and the Institute for Astronomy and Astrophysics of the University of T\"{u}bingen, with the support of DLR and the Max Planck Society. The Argelander Institute for Astronomy of the University of Bonn and the Ludwig Maximilians Universit\"{a}t Munich also participated in the science preparation for eROSITA.
\end{ack}

\end{document}